\documentclass[usenatbib]{mnras}

\usepackage{natbib}
\bibpunct{(}{)}{;}{a}{}{,}
\usepackage{graphicx}
\usepackage{txfonts}
\usepackage{xcolor}
\usepackage{multirow}
\usepackage{pst-plot}
\usepackage{breakurl}
\SpecialCoor
\usepackage{hyperref}

\def\xmm{{\em XMM-Newton}}

\def\chan{{\em Chandra}}
\def\asca{{\em ASCA}}
\def\beppo{{\em BeppoSAX}}
\def\rxte{{\em RXTE}}

\def\rosat{{\em ROSAT}}

\def\ginga{{\em Ginga}}

\def\xrism{{\em XRISM}}
\def\euve{{\em EUVE}}
\def\exosat{{\em EXOSAT}}
\def\athena{{\em Athena}}
\def\lem{{\em LEM}}

\definecolor{arancio}{rgb}{1,0.5,0}
\definecolor{viola}{rgb}{0.7,0,1}
\definecolor{verde}{rgb}{0.2,0.7,0.7}
\definecolor{cobalt}{rgb}{0.0, 0.28, 0.67}
\definecolor{airforceblue}{rgb}{0.36, 0.54, 0.66}
\definecolor{ballblue}{rgb}{0.13, 0.67, 0.8}
\definecolor{battleshipgrey}{rgb}{0.52, 0.52, 0.51}
\definecolor{darkgreen}{rgb}{0.0, 0.2, 0.13}

\title[\chan/HETG velocity measurements in stellar coronal sources]{\chan/HETG Doppler velocity measurements in stellar coronal sources}
\author[E. Bozzo et al.]{
E.\ Bozzo$^{1}$\thanks{E-mail: enrico.bozzo@unige.ch},
D. P.\ Huenemoerder$^{2}$,
M. Falanga$^{3,4}$,
S. Paltani$^{1}$,
E. Costantini$^{5,6}$,
J. de Plaa$^{5}$, 
L. Gu$^{5}$
\\
$^{1}$Department of Astronomy, University of Geneva, Chemin d'Ecogia 16, CH-1290 Versoix, Switzerland\\
$^{2}$Massachusetts Institute of Technology, Kavli Institute for Astrophysics and Space Research, 77 Massachusetts Ave., Cambridge, MA 02139, USA \\
$^{3}$International Space Science Institute (ISSI), Hallerstrasse 6, 3012 Bern, Switzerland \\
$^{4}$Physikalisches Institut, University of Bern, Sidlerstrasse 5, 3012 Bern, Switzerland\\
$^{5}$SRON Netherlands Institute for Space Research, Niels Bohrweg 4, 2333 CA Leiden, The Netherlands\\
$^{6}$Anton Pannekoek Institute, University of Amsterdam, Postbus 94249, NL-1090 GE Amsterdam, the Netherlands
}

\date{}

\begin{document}
\label{firstpage}
\pagerange{\pageref{firstpage}--\pageref{lastpage}}

\maketitle

\begin{abstract}
Stellar coronal sources have been observed in the past not only for their astrophysical interest in the field of binary system evolution and interaction, but also for their invaluable roles as benchmarks for plasma spectral models and as calibration sources for high resolution spectroscopic X-ray instruments. These include the gratings on-board \chan\ and \xmm,\ as well as the new generation of high resolution capable-detectors recently flown on-board \xrism\ and planned for the future also on-board the \athena\ and the \lem\ missions. In our previous paper exploiting \chan/HETG observations of the prototypical coronal source Capella, it has been shown that the centroid energies of the many X-ray emission lines detected in the spectrum of this object change as a function of time due to the Doppler modulation within the binary. This is an effect that needs to be corrected while performing calibrations of high resolution X-ray instruments. In this paper, we extend our previous work on Capella to other known stellar coronal sources which have been observed with the \chan/HETG (11 objects in total). We measure in several objects clear trends in the velocity shifts along the orbit of the primary star, meaning that in these sources one of the two star components is largely dominating the high energy emission. In a number of systems the trend in the velocity shift is not obvious. This can be ascribed to the fact that both stellar components contribute significantly to the X-ray emission.

\end{abstract}


\section{Introduction}
\label{sec:intro}

In the past decades, stellar coronal sources received substantial interest from different research communities thanks to their role of unique laboratories to study especially in the UV and X-ray domains photospheric, chromospheric and coronal activity, including star-spots and the dependence of dynamos on stellar spectral types and rotation. Literature papers have extensively investigated these systems to exploit their emission line-rich X-ray spectra to advance our understanding on plasma physics, as well as to understand highly debated aspects of 
binary stellar interactions especially in the Algol-like systems \citep[see, e.g.][for reviews]{peters01,guedel04, guedel06}. 

Beside all these topics, coronal stellar sources are also among the most exploited targets to perform calibrations of high energy space-based instruments, especially those focusing on high resolution spectroscopy in the X-ray domain (with a resolution down to a few eV in the 0.3-10~keV energy range). Among these, there are the reflection grating spectrometers (RGSs) on-board \xmm\ \citep{rgs} and both the low energy  \citep[LETG;][]{letg0} and high energy transmission gratings \citep[HETG;][]{hetg} on-board \chan.\  In 2023, the \xrism\ mission has brought to space the Resolve instrument, featuring a micro-calorimeter array which is able to achieve a resolution of $\lesssim$6~eV in the 0.3-12~keV energy range \citep{xrism1} and extend our capability of performing high resolution spectroscopic observations to fainter and fainter objects thanks to the larger effective area compared to the \chan\ and \xmm\ gratings \citep{xrismscience}. 
Stellar coronal sources have been observed regularly by \xmm\ and \chan\ to unveil/monitor trends in their wavelength/energy scale (i.e., the relation between the wavelength/energy of the detected photon and its most likely location on the detector), variations in the electronic readout gain of their CCD detectors (i.e., the conversion factor from the recorded electric charge to the corresponding readout pulse height), as well as the consistency over time of their dispersion relations, resolving powers, line response functions, and grating-to-detector alignment \citep{vries,letg,letg2,letg3,hetg}. Detailed studies of the X-ray emission lines in stellar coronal sources have thus uniquely provided on one side the possibility of testing accurately our knowledge of atomic transitions by fitting the spectra with more and more advanced physical plasma model \citep[see, e.g.,][and references therein]{gu22}, and on the other side to monitor on the time scale of decades the performance and stability of the available X-ray instrumentation in space. \xrism\ is also planning to regularly observe over the years of operations stellar coronal sources for calibration purposes \citep{xrismcal}. 

In a previous paper \citep[][hereafter Paper\,I]{bozzo23}, we exploited the database of the available \chan/HETG observations of the prototypical (and so far best known) stellar coronal source, Capella, to show that the centroid energies of its emission lines change over time due to Doppler modulation within the binary \citep[see also][]{ishibashi06}. In particular, the modulation appears to be clearly dominated by the cooler G8III giant in the system that is thus also confirmed to be the main emitter in the X-ray domain. The modulation was proven stable over a timescale of $\sim$22~years, leading to periodic shifts in the line centroids by up to $\sim\pm$50~km~s$^{-1}$ along the $\sim$104~d-long binary orbital period. 

Variations of the line centroids by tens of km~s$^{-1}$ due to Doppler effects have a significant impact on the outcome of the instrument calibrations and thus a proper knowledge of this phenomenon in different stellar coronal sources is needed in order to use  these celestial sources as calibration targets and maximize the correct exploitation of high resolution X-ray instruments down to their finest energy resolution\footnote{For a discussion of this effect, see \url{https://cxc.harvard.edu/newsletters/news_12/hetg.html}.} \citep[see, e.g.,][]{ishibashi04}. Orbital Doppler velocity shifts would be superimposed on any random short-term velocity variations (few tens of ks) which might be caused by intense flares from active stars, as those giving rise to coronal mass ejections \citep[see, e.g.,][and references therein]{argiroffi19}.  The latter are interesting, but here we will concentrate on the systematic and predictable orbital Doppler shifts, keeping in mind that flares could conceivably introduce some "background" noise.

In this paper, we extend our previous work on Capella to other well known stellar coronal sources, several of which are planned to be or could potentially be used as additional calibration targets for high resolution X-ray instruments. The availability of multiple suitable calibration targets in different regions of the sky and with different emission line properties is an advantage to ease observational plans and cross-check the outcome of the calibrations. We illustrate in Sect.~\ref{sec:sources} the selection method for the source candidates used in the present paper and provide a brief summary of their relevant Astrophysical properties. In Sect.~\ref{sec:analysis}, we summarize the data processing and analysis for all sources, including an overview of all obtained results (per source). We discuss all the main findings of our analyses in Sect.~\ref{sec:discussion}, providing an exhaustive examination of the best candidate stellar coronal sources for future calibration purposes. Our conclusions are reported in Sect.~\ref{sec:conclusion}.

\section{Source selection}
\label{sec:sources}

The goal of the present paper is to look for evidences of Doppler shifts in the centroid energies of emission lines in the spectra of stellar coronal sources by exploiting \chan/HETG observations and extend our previous analogous work which only considered the prototype of these systems, i.e. Capella (see Paper\,I). To select promising candidate sources for the study, we crossed-match the catalogue of chromospherically active binary stars \citep[CABS;][]{cabs} with the \chan\ grating-data archive and catalog \citep[TGCat;][]{tgcat}. We first screened the CABS including only sources with well determined orbital properties via radial velocity measurements and then verified which of these objects were observed with \chan\ ACIS-S in combination with the HETG. As extensively commented in Paper\,I, this combination of detectors and gratings in \chan\ provides the most accurately calibrated data-sets that can be exploited for the identification of velocity shifts in the range of few tens of km~s$^{-1}$ \citep[achieving a resolving power $\Delta E/E\gtrsim$1000; see also Sect.~\ref{sec:analysis} and][]{ayres01}. We ended-up with a total of 11 objects (excluding Capella), out of which 8 had a substantial orbital phase coverage with the available HETG data (HR\,1099, IM\,Peg, AR\,Lac, UX\,Ari, V824\,Ara, TZ\,CrB, HR\,5110,and $\sigma$\,Gem). For these 8 sources, we provide a brief description of their properties in the following sub-sections. For the three objects with limited coverage ($\lambda$\,And, II\,Peg,  and Ty\,Pyx), the data did not permit us to perform a thoughtful analysis and thus we report briefly about them only in the Appendix (see Sect.~\ref{sec:lambdaand}, \ref{sec:iipeg}, and \ref{sec:typyx}).

\subsection{HR\,1099}
\label{sec:hr1099_intro}

HR\,1099 (V711 Tau) comprises a K1\,IV and a G5\,V orbiting one another with a period of 2.8377~days. HR\,1099 is a member of the so-called RS CVn class characterized by a prominent magnetic activity which makes the source relatively bright across the electromagnetic spectrum \citep[see, e.g.,][]{donati99,strassmeier00}. \citet{frasca05} have shown that the source displays a strong variability in the orbital period ($P$), reporting an estimated $\Delta P/P\simeq 10^{-4}$ between 1980 and 2004. These authors suggested that the variation can be associated to the change of the gravitational quadrupole moment of the K1\,IV component. However, it can be appreciated in their Fig.~2 that the system ephemerides calculated originally by \citet{fekel83} cannot satisfactorily describe the radial velocity measurements acquired in the optical and ultraviolet domains at later stages (1994-2001). The variation in the orbital period is significant over the time scale of 2-3 decades and remains virtually undetected on the time scale of just a few years. For data in the period 1994-1997, updated ephemerides were provided by \citet{garcia03}. 

High resolution spectroscopy of HR\,1099 in the X-ray domain has been performed previously by \citet{audard01b} using the RGS on-board \xmm\ and by \citet{ayres01} using the \chan/HETG \citep[see also][for completeness]{drake01,uwe03,uwe04,nordon07}. The former authors showed that the source spectrum could be well characterized by using 4 distinct plasma components with the abundances of different elements linked among them. To fit the RGS spectra, they made use of the collisional ionization equilibrium CIE model available within the {\sc spex} software package 2.0 \citep{kaastra96}. As part of a multi-wavelength campaign aiming at studying flares from HR\,1099, \citet{ayres01} reported on an in-depth analysis of the different identified emission lines in the source HETG spectra. They commented in particular a  possible detection of a change in energy of the strongest Ne\,X $\lambda$12.1\AA\ line, qualitatively in agreement with the Doppler modulation expected for half of the system orbital period along the predicted radial velocity curve of the K1 star (see their Fig.~9a and 9b). \citet{ayres01} used in their work the 96~ks-long \chan/HETG ObsID.~62538, which was also reanalyzed later by \citet{huenemoerder13} in the context of a comparison between X-ray spectra at high resolution of stellar coronal sources and solar flares. However, these authors did not comment further on the energy variation of any of the detected lines from HR\,1099.

 \subsection{IM\,Peg}
 \label{sec:impeg_intro}

IM\,Pegasi (a.ka. HR\,8703, HD\,216489) is a binary system hosting a K2\,III giant (primary) and a K0\,V (secondary) in a circular orbit. The estimated orbital period is 24.65~days. The most updated system ephemerides were published by \citet{marsden05}. These authors exploited optical observations from the Automatic Spectroscopic Telescope (AST) of Tennessee State University to measure not only the radial velocity curve of the primary component but also for the first time that of the secondary component. They also improved the measurement of the source orbital period and constrained to an unprecedented accuracy the estimate of the mass ratio for the primary and secondary component \citep[see their Table~2 and see also][]{fekel99,berd99}. 

At the best of our knowledge, high resolution X-ray spectroscopy of IM\,Peg was only reported by \citet{westbrook08} using eight \chan/HETG observations performed between July and August 2002. These authors showed that the source HETG spectra could be well described by using two thermal plasma components with temperatures of about 1~keV and 3~keV, respectively (the fits to the spectra were performed within the {\sc Sherpa} environment; see their Sect.~3.4 and Table~5). The same dataset has been exploited also by \citet{testa04} and \citet{testa07} to analyze specific emission lines and investigate both the density and the X-ray optical depth of coronal plasmas \citep[see also][]{testa04b,ness04}.

 \subsection{AR\,Lac} 
 \label{sec:arlac_intro}

AR\,Lac comprises G2\,IV and K0\,IV sub-giants in a 1.98~d orbit \citep[see, e.g.,][for a review]{gehren99}. The system displays significant changes in the orbital period over decades, likely associated with its magnetic activity \citep[see, e.g.,][and references therein]{frasca00,frasca00b,siviero06,lu12}. The estimated distance to AR\,Lac is 42.5 pc \citep{gaia3}. The binary is known since the early 80's to be a relatively bright X-ray source in which both stellar components contribute to the high energy emission \citep{walter80,walter83,Huenemoerder13b}. \citet{white} estimated an absorption column density toward  AR\,Lac of 10$^{19}$~cm$^{-2}$ using data from the \asca\ observatory. 

AR\,Lac is also known to be an eclipsing binary, an aspect making this system particularly well suited to study and constrain the active region geometry. In this context, the most recent attempt to use multi-wavelength observations of AR\,Lac to disentangle the properties and variability over-time of the stellar coronae of its G2\,IV and K0\,IV components was presented by \citet{drake14}. These authors showed that the quiescent long term X-ray emission (excluding flares) of the system is remarkably stable but that primary eclipses (when the G2\,IV component lies behind the K0 star) display a hardly repeatable profile. Secondary eclipses are not observed in X-rays. The overall findings led the authors to conclude that AR\,Lac X-ray emission arises from both the stellar corona of the G2\,IV and K0\,IV star, as well as from surrounding extended structures \citep[see also][]{karakus21}. The more compact G2\,IV star is found to contribute a factor 2-5 more to the total X-ray emission compared to the K0\,IV.   
   
High resolution X-ray spectroscopy observations of AR\,Lac were presented in details by \citet{Huenemoerder13b}. These authors reported on the six \chan\ observations of the source available as of today with the ACIS-S used in combination with the HETG (they also made use of simultaneous UV data collected with the \euve\ satellite). The  analysis of individual X-ray emission lines allowed the authors to model the stellar coronae emission measures and abundance distributions, concluding that both the G2\,IV and the K0\,IV stars are contributing to the high energy emission and that there is a noticeable long-term stability in the overall coronal structure around the binary.

\subsection{UX\,Ari} 
\label{sec:uxari_intro}

UX\,Ari s a triple system, where the two main components constitute a RS CVn binary. These are classified as a K0 IV (primary) and a G5\,V star (secondary), respectively \citep[see, e.g.,][and references therein]{peterson11}. The estimated distance to the system is 50.5~pc. The triple nature of UX\,Ari is believed to be at the origin of the secular changes in the binary center of mass velocity, with values in the range 25.887-29.277~km~s$^{-1}$. A comprehensive overview of these measurements, as well as the most updated source ephemerides, have been reported by \citet{duemmler01}. Both the primary and the secondary component in UX\,Ari are long suspected to be chromospherically active, with the K0 \,IV star being largely dominant \citep[see, e.g.][]{huene89,ulvas03}. 

High resolution spectroscopy of the source in the X-ray domain has been carried out in the past by \citet{audard03}. These authors found that the source X-ray spectrum as observed by the RGS on-board \xmm\ could be well fit with a model comprising four CIE components with variable abundances for 11 elements (C, N, O, Ne, Mg, Si, S, Ar, Ca, Fe, and Ni). The only performed  \chan/HETG observation of UX\,Ari was preliminary reported by \citet{westbrook08} and exploited later by a number of authors for the study of a few specific lines aiming at investigating the size and density of coronal plasma, as well as the coronae X-ray optical depth \citep{ness03,ness04,testa04,testa07}.

\subsection{V824\,Ara} 
\label{sec:v824ara_intro}

V824\,Ara is a triple system consisting of an inner binary with a G5~IV and a K0~IV star, and an 
M3~V star that is spatially located 33\arcsec\ away from the inner binary system. 
The latter is characterized by a well known orbital period of 1.68~days, with the most recent ephemerides 
provided by \citet{strassmeier00b}. The distance to the system is measured at 30.5063~pc \citep{gaia18}.  

High resolution spectroscopy in the X-rays of this system was presented by \citet{lalitha15}, exploiting the only available \chan/HETG observation of UX\,Ari carried out on 12 January 2000. These authors showed that the HETG spectrum of the inner V824\,Ara system binary could be well described by using a 3 temperature thermal plasma emission model and that no major differences in the spectral parameters could be identified across the entire observation, although in the middle part of the \chan\ exposure and toward the last 20~ks flaring-like activities were observed. A similar spectral analysis could also be distinctly carried out for the M3 star component as this is well spatially separated by the \chan\ instruments. From the dispersed spectra, these authors derived emission measure distributions and element abundances for both the V824\,Ara inner binary and the M3 star, comparing values with those of other RS CVn type binary systems and active M dwarfs.

\subsection{TZ\,CrB} 
\label{sec:tzcrb_intro}

$\sigma^2$ CrB (TZ\,CrB) is part of a multiple system called $\sigma$ Coronae Borealis, located at a distance of 21.7~pc. TZ\,CrB is an active double-lined binary comprising a F9\,V primary and a G0\,V secondary, both close to the zero-age main sequence and relatively close in mass, stellar radius, effective temperature, and evolutionary status. The system measured orbital period is of about 1.14 day \citep[see][for an overview of the source]{strassmeier03}. Both stars in TZ\,CrB are known to be chromospherically active \citep[see, e.g.][and references therein]{osten03}, with the secondary usually displaying a slightly stronger activity than the primary \citep{frasca97}. The other components of the $\sigma$ Coronae Borealis hierarchical multiple system include $\sigma^1$ CrB, which is a solar-like star in visual binary system with TZ\,CrB (the estimated orbital period is about 900~years), and an M dwarf binary. $\sigma^1$ CrB is located 6.6\arcsec away from $\sigma^2$ CrB. The most recent ephemerides of TZ\,CrB have bee published by \citet{deepak09}. 

The X-ray spectrum of TZ\,CrB has been studied with multiple facilities in the past, including \ginga,\ \asca,\ and \rxte\ \citep{stern92, osten00, osten03}. The low resolution spectra of the source as observed by these instruments were mostly fit with a two temperature thermal plasma model (VMEKAL in {\sc xspec}), with temperatures of $\sim$0.7~keV and $\sim$2.0~keV, respectively. Analyses of high resolution X-ray spectra of TZ\,CrB were reported first by \citet{osten03} using \chan\ data collected with the ACIS-S in combination with the HETG and then also by \citet{suh05} using the RGS gratings on-board \xmm.\ The results presented by the former authors relied on a line-based analysis and were mostly focused on the determination of the plasma distribution with the temperature, as well as the elemental abundances. The latter authors adopted a global-fit technique for the RGS spectra of the source and showed that these could be satisfactorily fit with a model comprising four thermal plasma components. The temperatures of these components were of 0.33~keV, 0.67~keV, 1.27~keV, and 2.45~keV, respectively. \citet{suh05} discussed that the \xmm\ data did not show evidence for any X-ray emission from the nearby $\sigma^1$ CrB, but reported about a private communication from the authors of the 2003 paper \citep{osten03} explaining that the \chan\ data gave evidence of this source being much less luminous than the main target of all these studies, i.e. TZ\,CrB.

\subsection{HR\,5110} 
\label{sec:hr5110_intro}

HR\,5110 is an active binary system, hosting an F2\,IV primary star and and a K0\,IV secondary star in a 2.61~d-long circular orbit \citep[see, e.g.,][and references therein]{marenin86}. The system ephemerides have been published by \citet{eker87} and more recently (partly) revised by \citet{ransom03}. The distance to the systems is estimated at 46.888~pc \citep{gaia3}. 

In the X-ray domain, the source has been poorly studied. Low resolution spectra have been analyzed by \citet{graffagnino95} using \rosat\ data. These authors reported the sole measurements of the absorption column density in the direction of the source. This has been found to vary within the range $N_{\rm H}$$\sim$2-9$\times$10$^{19}$~cm$^{-2}$, depending also slightly on the model used to describe the continuum. High spectral resolution observations of the source have been carried out in the past with both the gratings on-board \chan\ and \xmm,\ but none of these data have yet been published (to the best of our knowledge).

\subsection{$\sigma$\,Gem} 
\label{sec:sigmagem_intro}

$\sigma$\,Gem is a chromospherically active binary that was for a long time classified as a single-lined spectroscopic system, as only the primary component was detected and classified as a K1 III star \citep[see, e.g.][and references therein]{henry95}. A major advancement in the understanding of this system was presented by \citet{rotten15}, who exploited interferometric infrared observations to detect for the first time the secondary star in the binary and improve our knowledge of the system geometry (the estimated distance to the source is of 38.8~pc). These authors published the most updated ephemerides of $\sigma$\,Gem and revised the classification of the primary star into a K giant. A definitive precise classification of the secondary component is still lacking but the same authors proposed that is is a main-sequence early K star.  

Observations of $\sigma$\,Gem in the X-ray domain with low energy resolution instruments were presented by \citet{singh87}. These authors exploited \exosat\ data to provide a description of the source emission below $\sim$10~keV and also provided an estimate of the average absorption column density in the direction of the binary (2.0$\times$10$^{19}$~cm$^{-2}$). At high energy resolution, the X-ray spectrum of $\sigma$\,Gem was studied with both data acquired by the \chan\ and \xmm\ gratings. \citet{huenemoerder13} presented an individual line-based analysis of the two \chan\ HETG observations of the source available so far, comparing the derived plasma properties with those from solar flares. The same dataset has been also exploited to investigate both the density and the X-ray optical depth of coronal plasmas \citep[see][]{testa04,testa04b,ness04,testa07}. \citet{nardone06} compared the quiescent \chan\ LETG spectrum of the source with that obtained from an \xmm\ grating observation carried out during a flare of the source, evidencing the appearance of a hotter plasma component during the enhanced X-ray activity period.

\section{Data processing and analysis}
\label{sec:analysis}

In continuation with what we have done in our Paper\,I, all data of the selected sources here were homogeneously processed with the {\sc chandra\_repro} tool (version 15 December 2022) available within the \chan\ data analysis software {\sc CIAO} v.4.14. Calibration files were obtained from our local installation of the \chan\ calibration database, CALDB, v.4.9.8. As all selected sources are relatively bright for the HETG, we followed the same data reduction steps described in Paper\,I for Capella. We run the {\sc chandra\_repro} script by setting the parameter {\sc tg\_zo\_position=detect} to correct pile-up issues and the parameter {\sc pix\_adj=NONE} to prevent randomization from impacting the outcome of the {\sc tg\_findzo} processing. For all data-sets, we made use of the $\pm$1 order spectra from both sets of gratings available within the HETG\footnote{See all details at \url{https://cxc.harvard.edu/proposer/POG/html/chap8.html}.}, i.e. the High Energy Grating (HEG) and the Medium Energy Grating (MEG). We limited our analysis to the energy interval 0.77-7.0~keV in order to have an homogeneous range for all MEG and HEG spectra (this is applied to all spectra and to all sources considered in this paper). We discuss specific analysis steps to the observations of the different sources in the following sub-sections. 
\begin{table*}
\caption{\chan/HETG observations of HR\,1099 considered in this work. For each observation, we report the ID, the middle observational time in heliocentric Julian day, the effective exposure time, the orbital phase calculated by using the ephemerides in \citet{garcia03}, and the velocity obtained from the \chan\ data using both the {\sc spex} and the {\sc xspec} software environments (corrected for the barycentric motion of the Earth around the center of mass of the solar system - this correction velocity is indicated explicitly in the rightmost column). All indicated uncertainties are at 1$\sigma$~c.l. The uncertainty on the phase includes the uncertainties on the ephemerides, as well as the duration of the observation. Note that ObsID~62538 was divided in 8 segments for the measurement of the Doppler velocities.} 
\label{tab:hr1099_obs}
\begin{tabular}{ccccccc}
\hline
 & Middle & Effective &   & Corrected velocity & Corrected velocity & Barycenter correction \\
 & Observational time & exposure & Phase & V$_{{\rm corr}_{\rm spex}}$ & V$_{{\rm corr}_{\rm xspec}}$ & V$_{\rm bary}$ \\
ObsID & (HJD)  & (ksec)   & $\phi$ & (km/s) & (km/s) & (km/s) \\
\hline
62538$_{\rm int1}$  & 2451436.5259 &  12.0 & 0.254$_{-0.036}^{+0.036}$ & -6.5$_{-3.9}^{+3.9}$   & 0.7$_{-3.8}^{+3.8}$ &  -24.2 \\ 
62538$_{\rm int2}$  & 2451436.6646 &  12.0 & 0.303$_{-0.036}^{+0.036}$ & -27.3$_{-3.6}^{+3.6}$   & -23.3$_{-3.7}^{+3.7}$ &  -24.2 \\ 
62538$_{\rm int3}$  & 2451436.8034 &  12.0 & 0.352$_{-0.036}^{+0.036}$ & -32.0$_{-3.5}^{+3.5}$   & -34.1$_{-3.5}^{+3.5}$ &  -24.1 \\ 
62538$_{\rm int4}$  & 2451436.9421 &  12.0 & 0.401$_{-0.036}^{+0.036}$ & -54.5$_{-3.6}^{+3.6}$   & -50.6$_{-3.4}^{+3.4}$ &  -24.1 \\ 
62538$_{\rm int5}$  & 2451437.0809 &  12.0 & 0.450$_{-0.036}^{+0.036}$ & -60.6$_{-4.2}^{+4.2}$   & -59.1$_{-3.5}^{+3.5}$ &  -24.0 \\ 
62538$_{\rm int6}$  & 2451437.2197 &  12.0 & 0.499$_{-0.036}^{+0.036}$ & -59.2$_{-3.5}^{+3.5}$   & -60.9$_{-3.3}^{+3.3}$ &  -24.0 \\ 
62538$_{\rm int7}$  & 2451437.3584 &  12.0 & 0.548$_{-0.036}^{+0.036}$ & -60.5$_{-3.6}^{+3.6}$   & -57.6$_{-3.4}^{+3.4}$ &  -23.9 \\ 
62538$_{\rm int8}$  & 2451437.4972 &  12.0 & 0.597$_{-0.036}^{+0.036}$ & -67.3$_{-3.6}^{+3.6}$   & -60.4$_{-3.5}^{+3.5}$ &  -23.9 \\ 
1252           & 2451439.1526 &  14.8 & 0.180$_{-0.036}^{+0.036}$ & 3.7$_{-3.8}^{+3.8}$   & 6.7$_{-3.7}^{+3.7}$ &  -23.5 \\ 
\hline
\end{tabular}
\end{table*}

\subsection{HR\,1099}
\label{sec:hr1099_data}

\chan\ observed HR\,1099 with the ACIS-S+HETG twice on 1999 September 14 at 22:53 (UTC) for a total exposure time of 96~ks (ObsID~62538) and on 1999 September 17 at 12:42 (UTC) for a total exposure time of 15~ks (ObsID~1252). According to \citet{ayres01}, the source was caught decaying from a flaring activity during the observation ObsID~62538 and decreasing toward a lower emission state in the ObsID~1252 (see their Fig.~1). To calculate the orbital phase of these observations, we used the ephemerides available in Table~5 of \citet{garcia03}, as these provide the closest orbital period parameter determinations to the \chan\ observations. Based on the discussion in Sect.~\ref{sec:hr1099_intro}, we considered negligible any variation in the system orbital properties occurred between the collection of the optical data in 1998 used by \citet{garcia03} and the epochs of the \chan\ observations. We defined as the zero orbital phase the point of maximum radial velocity for the K1\,III component, which is shifted in phase by 0.25 compared to the $T_0$ values provided in Table~5 of \citet{garcia03}. We also took into account in our phase determinations the additional shift of -0.0478 in phase reported by \citet{garcia03} and associated to the correction for the phase of the superior conjunction with respect to the ephemerides provided earlier in the literature by \citet{fekel83}.  
\begin{table}
    \begin{center}
    \caption{\label{tab:hr1099_spe} Spectral fit results obtained from the HEG +1 spectrum of HR\,1099 during the ObsID~62538. The best fit model comprises 4 CIE components plus a REDS component to take Doppler velocity effects into account. The abundances of relevant elements indicated in the table were left free to vary in the fit but the same abundances were linked across the different CIE components. The leftmost column gives the parameters of the models, while the two middle columns give the best determined value with the associated 1~$\sigma$~c.l uncertainty. The rightmost column reports the units of the different parameters (if any).  Note that the normalizations of the BVVAPEC components in {\sc xspec} are given in units of $(10^{-14}/4 \pi (D_{\rm A} (1+z))^{2}) \int n_{\rm e} n_{\rm H} dV$, where $D_{\rm A}$ is the angular size distance to the source (cm), while $n_{\rm e}$ and $n_{\rm H}$ are the electron and H densities (cm$^{-3}$)}. The third column from the left gives an overview of the results obtained from the fit to the HEG +1 spectrum in {\sc xspec}. $kT$ and $N$ represent the temperature and normalization of each CIE/BVVAPEC components, while the other symbols denote the abundances of the different elements considered. These are measured in units of proto-stellar abundances in the case of {\sc spex} \citep{lodders} and according to the tables in \citet{anders89} for {\sc xspec}. We also indicated the value of the velocity shift ($z$) and the value of the C-statistics with the number of degrees of freedom (d.o.f.).
    \begin{tabular}{llll}
        \hline
        \hline
        Model parameter & \multicolumn{1}{c}{Best fit values} & \multicolumn{1}{c}{Best fit values} & Units \\
                        & \multicolumn{1}{c}{{\sc spex}} & \multicolumn{1}{c}{{\sc xspec}} &  \\        
        \hline
        $kT_{1}$ & 0.11$_{-0.02}^{+0.02}$ & 0.100$_{-0.008}^{+0.014}$ & keV \\
        $kT_{2}$ & 0.52$_{-0.03}^{+0.06}$ & 0.55$_{-0.02}^{+0.02}$ & keV \\
        $kT_{3}$ & 2.52$_{-0.05}^{+0.04}$ & 2.54$_{-0.02}^{+0.02}$ & keV \\
        $kT_{4}$ & 1.03$_{-0.02}^{+0.02}$ & 1.06$_{-0.02}^{+0.02}$ & keV \\
        $N_{1}$ & 32.4$_{-13.3}^{+27.7}$ & 0.41$_{-0.25}^{+0.35}$ & 10$^{59}$~m$^{-3}$  \\
        $N_{2}$ & 1.5$_{-0.1}^{+0.1}$ & 0.014$_{-0.001}^{+0.001}$ & 10$^{59}$~m$^{-3}$  \\        
        $N_{3}$ & 7.6$_{-0.2}^{+0.2}$ & 0.068$_{-0.002}^{+0.002}$ & 10$^{59}$~m$^{-3}$  \\
        $N_{4}$ & 2.9$_{-0.1}^{+0.2}$ & 0.026$_{-0.001}^{+0.001}$ & 10$^{59}$~m$^{-3}$  \\
        Ne & 1.30$_{-0.05}^{+0.04}$ & 1.46$_{-0.06}^{+0.06}$ & \\
        Na & 1.2$_{-0.4}^{+0.4}$ & 0.6$_{-0.3}^{+0.3}$ & \\
        Mg & 0.36$_{-0.02}^{+0.02}$ & 0.36$_{-0.02}^{+0.02}$ & \\
        Al & 0.6$_{-0.1}^{+0.1}$ & 0.6$_{-0.1}^{+0.1}$ & \\
        Si & 0.31$_{-0.02}^{+0.02}$ & 0.32$_{-0.02}^{+0.02}$ & \\
        S & 0.28$_{-0.03}^{+0.03}$ & 0.28$_{-0.03}^{+0.03}$ & \\
        Ca & 0.5$_{-0.1}^{+0.1}$ & 0.4$_{-0.1}^{+0.1}$ & \\
        Fe & 0.26$_{-0.01}^{+0.01}$ & 0.178$_{-0.007}^{+0.007}$ & \\
        $z$ & -71.3$_{-2.9}^{+2.8}$ & -72.0$_{-2.7}^{+2.8}$ & km~s$^{-1}$ \\    
        C-statistics/d.o.f. & 2683.1/2038 & 3099.7/2393 & \\
        \hline
        \hline
    \end{tabular}
\end{center}
\end{table}

To obtain reliable measurements of the Doppler velocities of the emission lines in the source HETG spectra, we first extracted the $\pm$1 order HEG and MEG spectra from the ObsID~62538, which is endowed with the longest exposure and highest statistics. Following previous results in the literature, we first fit the HEG +1 spectrum with a model comprising 4 CIE components available within the {\sc spex} fitting software package\footnote{\url{https://zenodo.org/doi/10.5281/zenodo.1924563}.} version 3.08.00 with the atomic database version 3.08.00  \citep{kaastra96,spex30702}. We left the abundances of the different elements already identified by \citet{huenemoerder13} free to vary in the fit but linked together the abundances of the 4 CIE components. A REDS component (i.e., a multiplicative model applying a redshift $z$ to an arbitrary additive component) was also included to take into account the effect of the Doppler velocity shifts within the binary. We verified that the addition of a HOT component (i.e., a collisional ionisation equilibrium absorption model) to mimic the effect of the Galactic absorption does not have a measurable impact on the best measured spectral parameters due to the known low value of the column density in the direction of the source \citep[see, e.g.][]{piskunov97}. We also verified that leaving the velocity broadening parameters of the 4 CIE components free to vary in the fit did not lead to any significant improvement of the results, neither affected significantly the temperatures of the four components and the Doppler velocity values. We thus fixed for all fits the broadening velocity to zero (see also Paper\,I). The distance to the source in {\sc spex} was fixed at 29.63~pc \citep{gaia2}. Although the goal of the present paper is to provide Doppler velocity measurements and not to achieve an exhaustive plasma modeling for the different stellar coronal sources, we summarize in Table~\ref{tab:hr1099_spe} for completeness the best parameter values obtained from the fit to the HEG +1 spectrum of the observation ObsID~62538. This spectrum, together with the best fit model and the residuals, is shown in Fig.~\ref{fig:hr1099_plot_spex}. Note that the values obtained are qualitatively similar (in terms of temperatures of the plasma components and element abundances) to those reported in previous literature papers \citep[see, e.g., Table~3 in][]{audard01b}. We then fit all the other remaining spectra from the same observation (MEG +1, MEG -1, and HEG -1) one by one with the same model but fixing the temperatures of the 4 CIE components and the abundances to the values determined from the HEG +1 spectrum (we verified \emph{a posteriori} that no significant improvement in the fits could be obtained by leaving these parameters free to vary). Individual fits to all spectra allowed us to visually inspect each time the residuals and verify that all identified emission lines are carefully taken care of within the model. 
\begin{figure}
  \centering
  \includegraphics[width=9.0cm]{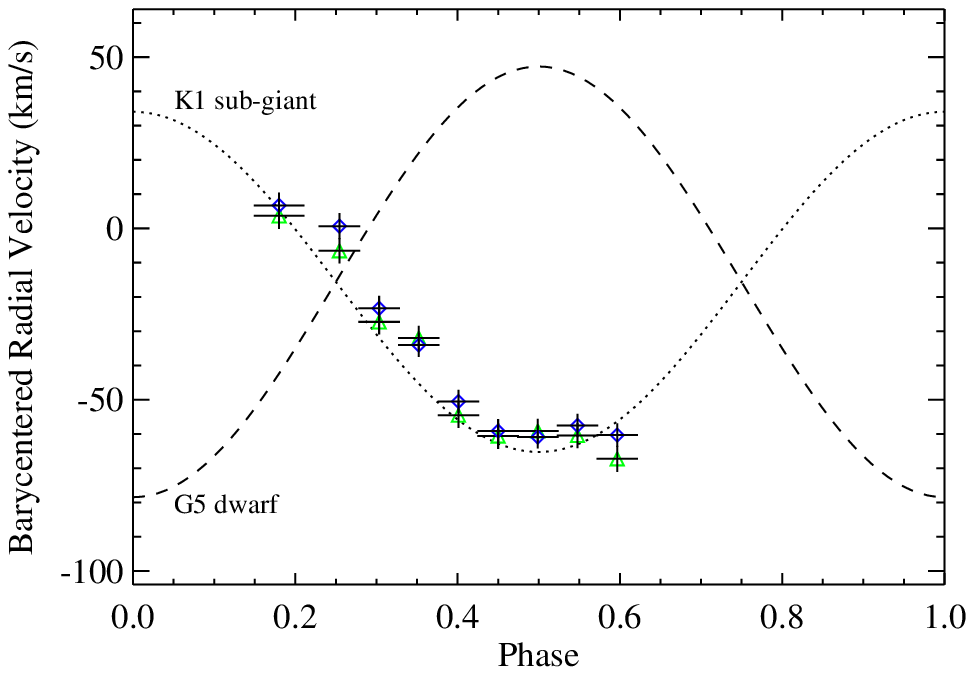}
  \caption{\label{fig:hr1099_orbit} Barycentered radial velocity measurements obtained from the \chan/HETG observations of HR\,1099 (see Table~\ref{tab:hr1099_obs}). The dotted and dashed lines are the expected barycentered radial velocities of the K1 sub-giant and G5 dwarf stars (see Sect.~\ref{sec:intro}) calculated according to the ephemerides published by \citet{garcia03}. The green points correspond to values obtained from the spectral fits in {\sc spex}, while blue points were obtained from the fits in {\sc xspec}. The error bars for each radial velocity measurement are given at 1~$\sigma$~c.l. and are only statistical. The uncertainty on the phase for each measurement includes the duration of the observation, as well as the uncertainties associated to the ephemerides.} 
\end{figure}

To investigate the effect of the Doppler modulation within the binary, we divided the ObsID~62538 in 8 segments covering each about 12~ks of effective exposure. This allowed us to retain for all spectra an adequate S/N that is also similar to that available for the ObsID~1252. For each segment, we extracted the four $\pm$1 order HEG and MEG spectra, building the corresponding arf and rmf files via the {\sc tgextract} and {\sc mktgresp} tools available within the {\sc CIAO} software. Each of the four spectra within each segment was fit with the best model described above, keeping the temperatures and abundances of the CIE components fixed and leaving only their normalizations plus the velocity shift parameter free to vary in the fits. In all cases, we obtained fully acceptable results with no evident structures in the residuals from the fits that could indicate any major discrepancy with respect to the observed emission lines and/or the need of additional plasma components. We obtained the Doppler shift measurement in each segment by performing a weighted average of the measurements obtained in each of the four $\pm$1 order HEG and MEG spectra. The same analysis technique was also applied to the observation ObsID~1252, which was not split in segments as the total exposure time was of only 14.8~ks. The same spectral model used for the ObsID~62538 could also satisfactorily fit the data in ObsID~1252, when the normalizations of the 4 CIE components and the velocity shift value were left free to vary in the fit. We verified that leaving the temperatures of the 4 CIE components as well as the abundances of one CIE component free to vary did not produce significant changes compared to the fit with these values fixed to those measured from the HEG +1 spectrum in ObsID~62538. All measured Doppler velocities from the \chan/HETG observations of HR\,1099, together with the estimated orbital phases, are reported in Table~\ref{tab:hr1099_obs} and plot in Fig.~\ref{fig:hr1099_orbit}. All Doppler measurements ($V_{\rm app}$) were corrected for the barycentric motion of the Earth around the center of mass of the solar system ($V_{\rm corr}=V_{\rm app}-V_{\rm bary}$, see Table~\ref{tab:hr1099_obs} and Paper\,I). The correction was derived using the algorithm developed by \citet{stumpff80} as implemented in the {\sc ispec} software \citep{ispec1,ispec2}. 

Looking at Fig.~\ref{fig:hr1099_orbit}, we note that the Doppler velocities measured from the \chan\ data are in good agreement with the predicted values along the curve of the K1 sub-giant (also considering that the error bars in the plot are at 1~$\sigma$~c.l.). To further consolidate these results, we re-performed the fits of all spectra changing from the {\sc spex} to the {\sc xspec} environment \citep[v.12.12.1;][]{xspec}. We used for all fits the same model as described above, exploiting in {\sc xspec} the BVVAPEC\footnote{For all details see \url{https://heasarc.gsfc.nasa.gov/xanadu/xspec/manual/node136.html}.} implementation for the plasma components (this component includes also a parameter accounting for the Doppler velocity shift). Although this is virtually equivalent to the {\sc CIE} component in {\sc spex}, the two environments are using different atomic databases (AtomDB v.3.0.9 was used for {\sc xspec}). The fits in {\sc xspec} were performed by leaving the temperatures and the element abundances for the plasma components free to vary in the initial fit to the HEG +1 spectrum (see Table~\ref{tab:hr1099_spe}), but we then fixed the abundances in all remaining fits (as done for the {\sc spex} analysis; see also Fig.~\ref{fig:hr1099_plot_spex}). The results obtained from the fits within {\sc xspec} are also plot along those obtained from {\sc spex} in Fig.~\ref{fig:hr1099_orbit}. All measurements are consistent to within the 1~$\sigma$~c.l. associated uncertainties.

 \subsection{IM\,Peg}
 \label{sec:impeg_data}

We re-analyzed in the case of IM\,Peg all the eight HETG observations that were previously reported by \citet{westbrook08}. During all these exposures, the source displayed a fairly stable X-ray emission \citep{huenemoerder02}. We adopted the same processing and fitting strategy illustrated in Sect.~\ref{sec:hr1099_data} for the source HR\,1099. For each of the eight observations, we extracted the average $\pm$1 order HEG and MEG spectra with the {\sc chandra\_repro} tool. Following previous results in the literature (see Sect.~\ref{sec:impeg_intro}), the order -1 HEG spectrum in the ObsID~2527 was fit within {\sc spex} by adopting a model comprising two CIE components with temperatures, element abundances, and normalizations left free to vary in the fit. We also included a REDS component to take into account the Doppler effect within the binary on the line centroid energies. The model provided a fully acceptable description of the data. We checked that the inclusion in the fit of a HOT component to mimic the Galactic absorption in the direction to the source did not significantly affect the fit parameters (to within the reported uncertainties at 1~$\sigma$~c.l.). The column density of this component was fixed in the fit to the value of 7.6$\times$10$^{20}$~cm$^{-2}$, as obtained from the {\sc Heasarc} on-line tool\footnote{\url{https://heasarc.gsfc.nasa.gov/cgi-bin/Tools/w3nh/w3nh.pl?}}. We note that this is an upper limit to the column density in the direction of the source, as the on-line tool returns the integrated value across our entire Galaxy, while the estimated distance to IM\,Peg is of only 96.4~pc \citep[this is also the distance we used in our {\sc spex} fits;][]{ratner12}. At the best of our knowledge, no more accurately determined values of the absorption column density in the direction of IM\,Peg were reported in the literature\footnote{Note that a tentative estimate of the absorption column density was obtained from \citet{westbrook08} during the fits to the HETG spectra. These values range from 7$\times$10$^{19}$~cm$^{-2}$ to 8.5$\times$10$^{20}$~cm$^{-2}$, being thus in agreement with our conclusion that the tested value of 7.6$\times$10$^{20}$~cm$^{-2}$ is an upper limit (yet having no significant impact on the measured  spectral parameters of interest for this paper).}. For completeness, we summarize the best fit values of the model parameters determined from the HEG -1 spectrum in Table~\ref{tab:impeg_spe} (see also Fig.~\ref{fig:implot_plot_spex} for a graphical representation of the source HEG -1 spectrum, together with the best fit model and the residuals from the fit). 
\begin{table}
    \begin{center}
 \caption{\label{tab:impeg_spe} Same as Table~\ref{tab:hr1099_spe}, but for the case of the -1 order HEG spectrum of IM\,Peg obtained from the ObsID~2527. The best fit was obtained by using two plasma components (CIE in {\sc spex} and BVVAPEC in {\sc xspec}). The fit in {\sc spex} also includes a REDS component.}
    \begin{tabular}{llll}
        \hline
        \hline
        Model parameter & \multicolumn{1}{c}{Best fit values} & \multicolumn{1}{c}{Best fit values} & Units \\
                        & \multicolumn{1}{c}{{\sc spex}} & \multicolumn{1}{c}{{\sc xspec}} &  \\        
        \hline
        $kT_{1}$ & 2.4$_{-0.1}^{+0.1}$ & 2.6$_{-0.1}^{+0.1}$ & keV \\
        $kT_{2}$ & 0.89$_{-0.04}^{+0.04}$ & 0.94$_{-0.04}^{+0.04}$ & keV \\
        $N_{1}$ & 2.4$_{-0.1}^{+0.1}$ & 0.0192$_{-0.0009}^{+0.0009}$ & 10$^{60}$~m$^{-3}$  \\
        $N_{2}$ & 0.69$_{-0.09}^{+0.10}$ & 0.0053$_{-0.0006}^{+0.0007}$ & 10$^{60}$~m$^{-3}$  \\        
        Ne & 1.7$_{-0.2}^{+0.2}$ & 2.1$_{-0.3}^{+0.3}$ & \\
        Na & 2.3$_{-1.3}^{+1.7}$ & 2.3$_{-1.6}^{+2.0}$ & \\
        Mg & 0.43$_{-0.08}^{+0.09}$ & 0.5$_{-0.1}^{+0.1}$ & \\
        Al & 0.3$_{-0.3}^{+0.4}$ & 0.3$_{-0.3}^{+0.5}$ & \\
        Si & 0.42$_{-0.07}^{+0.07}$ & 0.47$_{-0.08}^{+0.09}$ & \\
        S & 0.4$_{-0.1}^{+0.1}$ & 0.5$_{-0.2}^{+0.2}$ & \\
        Ca & 1.1$_{-0.5}^{+0.6}$ & 1.0$_{-0.5}^{+0.6}$ & \\
        Fe & 0.45$_{-0.05}^{+0.06}$ & 0.31$_{-0.04}^{+0.04}$ & \\
        $z$ & 12.9$_{-11.4}^{+11.4}$ & -3.0$_{-10.5}^{+10.5}$ & km~s$^{-1}$ \\    
        C-statistics/d.o.f. & 1963.7/1648 & 1826.3/1566 & \\
        \hline
        \hline
    \end{tabular}
\end{center}
\end{table}

As done in the case of HR\,1099, we then applied the same spectral model described above to fit the remaining 3 spectra in the ObsID~2527, as well as all other spectra obtained from the ObsID~2528-2534. The fits to all these spectra were carried out with the abundances of the different elements and the temperatures of the two CIE components fixed to the values reported in Table~\ref{tab:impeg_spe}. Only the normalizations of the CIE components and the velocity shift value were left free to vary in the fits (we obtained in all cases satisfactorily description of the data with no evident emission line left unfit). The value of the velocity shift in each observation was obtained as a weighted average of the values obtained from the corresponding $\pm$1 HEG and MEG spectra (four velocity shifts averaged for each observation). All fits were then repeated in the {\sc xspec} environment as done for HR\,1099. 
\begin{figure}
  \centering
  \includegraphics[width=9.0cm]{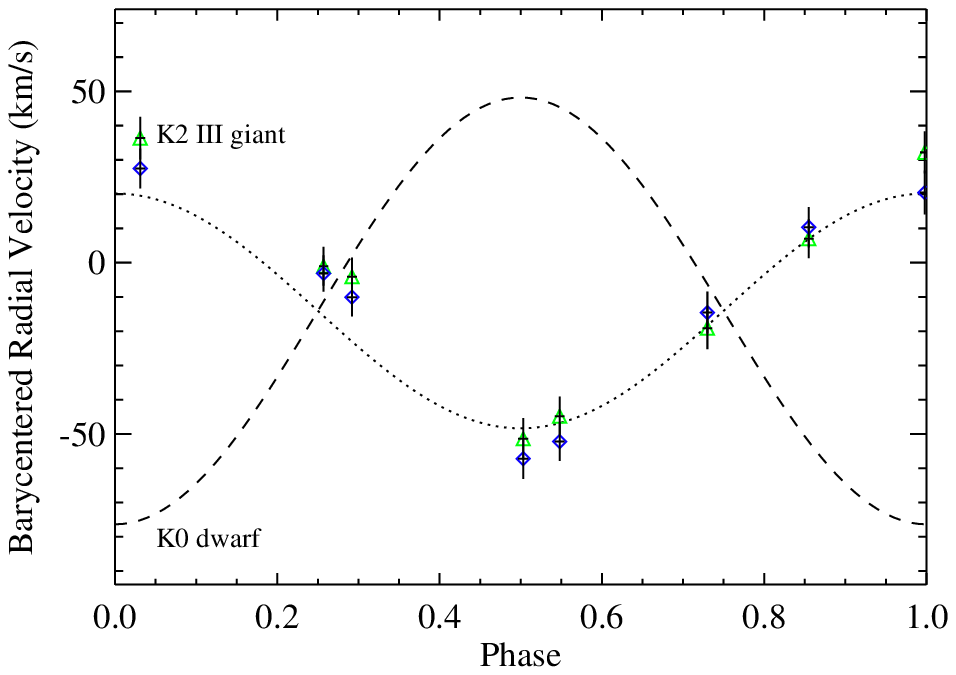}
  \caption{\label{fig:impeg_orbit} Same as Fig.~\ref{fig:hr1099_orbit} but for the case of IM\,Peg. The dotted and dashed lines are the expected barycentered radial velocities of the K2 III giant and K0 dwarf stars (see Sect.~\ref{sec:intro}) calculated according to the ephemerides published by \citet{marsden05}. The green points correspond to values obtained from the spectral fits in {\sc spex}, while blue points were obtained from the fits in {\sc xspec}. The error bars for each radial velocity measurement are given at 1~$\sigma$~c.l. and are only statistical. The uncertainty on the phase for each measurement includes the duration of the observation, as well as the uncertainties associated to the ephemerides.} 
\end{figure}

We then calculated the orbital phase of each HETG observation by using the ephemerides from \citet{marsden05}. As their time for the phase 0 is provided at conjunction, we shifted the phases for our plot by 0.75 (see their Fig.~2 and Table~2). We also took into account their shift in the overall velocities of -14.09~km~s$^{-1}$ to match the previously reported ephemerides by \citet{fekel99} and \citet{berd99}. We included in our estimate of the orbital phase of each \chan\ observation the duration of the exposure, as well as the uncertainties indicated by \citet{marsden05} on the estimated reference time and the source orbital period. The barycentric correction on all velocities was then included as done previously for HR\,1099. All results obtained from the \chan/HETG data of IM\,Peg are summarized in Table~\ref{tab:impeg_obs}. The results in this table are also shown in Fig.~\ref{fig:impeg_orbit}, where we observe that all measurements seem to follow rather accurately the velocities expected along the orbit of the K2\,III giant star. 
Measurements obtained from {\sc spex} and {\sc xspec} are consistent to within the 1~$\sigma$~c.l. associated uncertainties. 
\begin{table*}
\caption{Same as Table~\ref{tab:hr1099_obs} but for the \chan/HETG observations of IM\,Peg. The orbital phases were calculated from the ephemerides published by \citet{marsden05}. All indicated uncertainties are at 1$\sigma$~c.l.} 
\label{tab:impeg_obs}
\begin{tabular}{ccccccc}
\hline
 & Middle & Effective &   & Corrected velocity & Corrected velocity & Barycenter correction \\
 & Observational time & exposure & Phase & V$_{{\rm corr}_{\rm spex}}$ & V$_{{\rm corr}_{\rm xspec}}$ & V$_{\rm bary}$ \\
ObsID & (HJD)  & (ksec)   & $\phi$ & (km/s) & (km/s) & (km/s) \\
\hline
2527  & 2452457.3112 &  24.64 & 0.031$_{-0.006}^{+0.006}$ & 36.36$_{-6.2}^{+6.2}$   & 27.5$_{-5.8}^{+5.8}$ &  -25.8 \\ 
2528  & 2452463.7457 &  24.83 & 0.292$_{-0.006}^{+0.006}$ & -4.1$_{-5.6}^{+5.6}$   & -10.1$_{-5.6}^{+5.6}$ &  -24.73 \\ 
2529  & 2452468.9314 &  24.83 & 0.503$_{-0.006}^{+0.006}$ & -51.4$_{-6.0}^{+6.0}$   & -57.2$_{-6.0}^{+6.0}$ &  -23.65 \\ 
2530  & 2452474.5371 &  23.86 & 0.730$_{-0.006}^{+0.006}$ & -19.1$_{-6.1}^{+6.1}$   & -14.6$_{-6.1}^{+6.1}$ &  -22.27 \\ 
2531  & 2452481.1392 &  23.87 & 0.998$_{-0.006}^{+0.006}$ & 32.2$_{-6.1}^{+6.1}$   & 20.5$_{-6.4}^{+6.4}$ &  -20.41 \\ 
2532  & 2452487.5121 &  22.47 & 0.257$_{-0.006}^{+0.006}$ & -1.0$_{-5.7}^{+5.7}$   & -3.1$_{-5.3}^{+5.3}$ &  -18.38 \\ 
2533  & 2452494.6895 &  23.70 & 0.548$_{-0.006}^{+0.006}$ & -44.8$_{-5.7}^{+5.7}$   & -52.3$_{-5.6}^{+5.6}$ &  -15.82 \\ 
2534  & 2452502.2576 &  23.89 & 0.855$_{-0.006}^{+0.006}$ & -7.0$_{-5.7}^{+5.7}$   & 10.3$_{-5.9}^{+5.9}$ &  -12.85 \\ 
\hline
\end{tabular}
\end{table*}

 \subsection{AR\,Lac} 
 \label{sec:arlac_data}

We re-analyzed the six ACIS-S/HETG observations that were previously reported by \citet{Huenemoerder13b}. Four of these observations (ObsID~7,8,10, and 11) were carried out during the known system primary eclipses, while two longer exposures were obtained during quadrature (ObsID~6 and 9). The sourced displayed a remarkable X-ray variability across these observations, with a flare detected during the ObsID~7 \citep{Huenemoerder13b}. The details of these observations, including the estimated orbital phase, are reported in Table~\ref{tab:arlac_obs}. For the orbital phases, we made use of the ephemerides published by \citet{marino98}, as these are the closest to the \chan\ observations and minimize the effect of the source orbital period change (see Sect.~\ref{sec:arlac_intro}). In the case of AR\,Lac, the uncertainty we reported on the estimated orbital phases for the different observations include only the effect of the observational extension in time as no explicit uncertainties were indicated by \citet{marino98} on the determined reference time and system orbital period. 
\begin{table*}
\caption{Same as Table~\ref{tab:hr1099_obs} but for the \chan/HETG observations of AR\,Lac. The orbital phases were calculated from the ephemerides published by \citet{marino98}. All indicated uncertainties are at 1$\sigma$~c.l. Note that the ObsID~6 and 9 were divided into 4 segments in order to more closely follow any evolution of the velocity shifts as a function of the orbital phase (see text for more details).} 
\label{tab:arlac_obs}
\begin{tabular}{ccccccc}
\hline
 & Middle & Effective &   & Corrected velocity & Corrected velocity & Barycenter correction \\
 & Observational time & exposure & Phase & V$_{{\rm corr}_{\rm spex}}$ & V$_{{\rm corr}_{\rm xspec}}$ & V$_{\rm bary}$ \\
ObsID & (HJD)  & (ksec)   & $\phi$ & (km/s) & (km/s) & (km/s) \\
\hline
6$_{\rm int1}$  & 2451799.4659 &  7.9 & 0.916$_{-0.023}^{+0.023}$ & -30.5$_{-8.0}^{+8.0}$ & -37.0$_{-8.1}^{+8.1}$ &  -2.7 \\ 
6$_{\rm int2}$  & 2451799.5619 &  7.9 & 0.965$_{-0.023}^{+0.023}$ &  -23.4$_{-7.7}^{+7.7}$ & -29.4$_{-7.1}^{+7.1}$ &  -2.6 \\ 
6$_{\rm int3}$ & 2451799.6545 &  7.9 & 0.011$_{-0.023}^{+0.023}$ & -32.8$_{-8.0}^{+8.0}$ & -49.6$_{-7.7}^{+7.7}$ &  -2.6 \\ 
6$_{\rm int4}$  & 2451799.76304 & 8.4 & 0.066$_{-0.025}^{+0.025}$ & -33.1$_{-6.9}^{+6.9}$ & -45.6$_{-8.1}^{+8.1}$ &  -2.6 \\ 
7  & 2451804.0494 & 7.4 & 0.227$_{-0.022}^{+0.022}$ & -22.0$_{-5.8}^{+5.8}$ & -37.7$_{-5.9}^{+5.9}$ &  -1.2 \\ 
8  & 2451803.1516 & 7.4 & 0.775$_{-0.022}^{+0.022}$ & -10.5$_{-6.2}^{+6.2}$ & -17.9$_{-6.1}^{+6.1}$ &  -1.5 \\ 
9$_{\rm int1}$  & 2451805.3840 &  7.9 & 0.900$_{-0.023}^{+0.023}$ & -7.6$_{-7.2}^{+7.2}$ & -19.5$_{-7.3}^{+7.3}$ &  -0.8 \\ 
9$_{\rm int2}$  & 2451805.4766 &  7.9 & 0.947$_{-0.023}^{+0.023}$ &  19.5$_{-6.7}^{+6.7}$ & 6.6$_{-7.5}^{+7.5}$ &  -0.8 \\ 
9$_{\rm int3}$ & 2451805.5692 &  7.9 & 0.994$_{-0.023}^{+0.023}$ & 7.4$_{-7.4}^{+7.4}$ & -0.2$_{-7.1}^{+7.1}$ &  -0.8 \\ 
9$_{\rm int4}$  & 2451805.6875 & 8.5 & 0.053$_{-0.025}^{+0.025}$ & 3.7$_{-7.4}^{+7.4}$ & -10.2$_{-7.7}^{+7.7}$ &  -0.7 \\ 
10  & 2451807.9513 & 7.2 & 0.195$_{-0.021}^{+0.021}$ & -2.8$_{-7.5}^{+7.5}$ & -12.6$_{-7.7}^{+7.7}$ &  0.0 \\ 
11  & 2451807.1646 & 7.2 & 0.798$_{-0.021}^{+0.021}$ & -37.5$_{-8.2}^{+8.2}$ & -29.8$_{-7.7}^{+7.7}$ &  -0.3 \\ 
\hline
\end{tabular}
\end{table*}
\begin{table}
    \begin{center}
 \caption{\label{tab:arlac_spe} Same as Table~\ref{tab:hr1099_spe}, but for the case of the +1 order HEG spectrum of AR\,Lac obtained from the ObsID~6. The best fit was obtained by using two plasma components (CIE in {\sc spex} and BVVAPEC in {\sc xspec}). The fit in {\sc spex} also includes a REDS component.}
    \begin{tabular}{llll}
        \hline
        \hline
        Model parameter & \multicolumn{1}{c}{Best fit values} & \multicolumn{1}{c}{Best fit values} & Units \\
                        & \multicolumn{1}{c}{{\sc spex}} & \multicolumn{1}{c}{{\sc xspec}} &  \\        
        \hline
        $kT_{1}$ & 1.62$_{-0.08}^{+0.08}$ & 1.78$_{-0.07}^{+0.10}$ & keV \\
        $kT_{2}$ & 0.79$_{-0.03}^{+0.03}$ & 0.84$_{-0.02}^{+0.02}$ & keV \\
        $N_{1}$ & 3.6$_{-0.3}^{+0.4}$ & 0.015$_{-0.001}^{+0.001}$ & 10$^{59}$~m$^{-3}$  \\
        $N_{2}$ & 2.4$_{-0.3}^{+0.3}$ & 0.011$_{-0.001}^{+0.001}$ & 10$^{60}$~m$^{-3}$  \\
        O & 1.2$_{-0.5}^{+0.6}$ & 1.0$_{-0.4}^{+0.5}$ & \\        
        Ne & 1.5$_{-0.2}^{+0.2}$ & 1.9$_{-0.2}^{+0.3}$ & \\
        Na & 1.4$_{-0.9}^{+1.2}$ & 0.9$_{-0.8}^{+1.1}$ & \\
        Mg & 0.46$_{-0.06}^{+0.07}$ & 0.49$_{-0.06}^{+0.07}$ & \\
        Al & 0.5$_{-0.2}^{+0.3}$ & 0.5$_{-0.3}^{+0.3}$ & \\
        Si & 0.34$_{-0.05}^{+0.05}$ & 0.37$_{-0.05}^{+0.06}$ & \\
        S & 0.3$_{-0.1}^{+0.1}$ & 0.3$_{-0.1}^{+0.1}$ & \\
        Ar & 0.5$_{-0.3}^{+0.3}$ & 0.6$_{-0.3}^{+0.3}$ & \\
        Ca & 0.8$_{-0.5}^{+0.6}$ & 0.7$_{-0.4}^{+0.5}$ & \\
        Fe & 0.38$_{-0.04}^{+0.05}$ & 0.28$_{-0.03}^{+0.03}$ & \\
        $z$ & -32.7$_{-9.0}^{+9.0}$ & -16.7$_{-8.7}^{+8.6}$ & km~s$^{-1}$ \\    
        C-statistics/d.o.f. & 2122.6/1670 & 2046.9/1596 & \\
        \hline
        \hline
    \end{tabular}
\end{center}
\end{table}

Following the previous results in the literature and the analysis techniques described in the past sections for the other sources, we first extracted the HEG +1 AR\,Lac HETG spectrum from the ObsID~6, using the whole exposure time available. This spectrum was fit within {\sc spex} with a model comprising two CIE components, a REDS component to evaluate the binary velocity shifts, and a HOT component to mimic the effect of the Galactic absorption. We verified that the HOT component did not lead to significant changes in the spectral parameters of the CIE components, given the low value of the known column density in the direction of AR\,Lac (see Sect.~\ref{sec:arlac_intro}). In the fit, we left free to vary the abundances of the relevant elements whose emission lines were already identified in the HETG spectra of the source \citep{Huenemoerder13b}. The best fit parameters obtained from the HEG +1 spectrum are reported in Table~\ref{tab:arlac_spe} (see also Fig.~\ref{fig:arlac_plot_spex}). We then used this best determined spectral model to fit the data obtained from the $\pm$1 order MEG and HEG spectra of all other observations, keeping only the normalizations of the two CIE components and the velocity shift free to vary. In all cases, we obtained satisfactorily fits, with no evidence of emission lines being poorly described in the residuals from the fits. As done before, the velocity shift in each observation was obtained as the weighted average of the 4 values obtained from the fits to the corresponding $\pm$1 order HEG and MEG spectra. Given that the statistics of the ObsID~7, 8, 10, and 11 proved sufficient to obtain accurate measurements of the binary velocity shifts, we divided the ObsID~6 and 9 in 4 segments with an exposure time of roughly 8~ks in order to obtain a more detailed view of the velocity shift variations as a function of the orbital phase. 

\begin{figure}
  \centering
  \includegraphics[width=9.0cm]{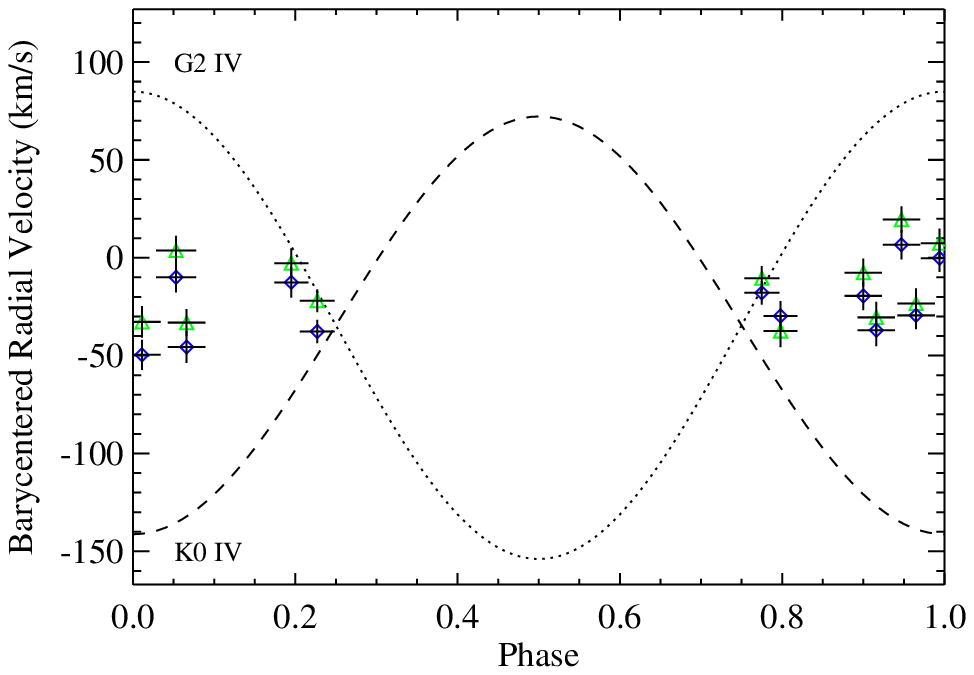}
  \caption{\label{fig:arlac_orbit} Same as Fig.~\ref{fig:hr1099_orbit} but for the case of AR\,Lac. The dotted and dashed lines are the expected barycentered radial velocities of the G2\,IV and K0\,IV stars (see Sect.~\ref{sec:intro}), calculated according to the ephemerides published by \citet{marino98}. The green points correspond to values obtained from the spectral fits in {\sc spex}, while blue points were obtained from the fits in {\sc xspec}. The error bars for each radial velocity measurement are given at 1~$\sigma$~c.l. and are only statistical. The uncertainty on the phase for each measurement includes only the duration of the observation (see text for details).} 
\end{figure}

In order to continue the comparison initiated for HR\,1099, we fit also also for the case of AR\,Lac all spectra by using the {\sc xspec} environment. We first fit the ObsID~6 +1 HEG spectrum with a model comprising two BVVAPEC components and then used this best model to describe all other spectra as well (leaving only the normalization of the two components and their linked velocity shifts to vary in the fits). The results of all these analyses are summarized in Table~~\ref{tab:arlac_obs} and plot in Fig.~\ref{fig:arlac_orbit}. Measurements obtained from {\sc spex} and {\sc xspec} are all consistent to within $\sim$1~$\sigma$~c.l. associated uncertainties. 

It is interesting to note from Fig.~\ref{fig:arlac_orbit} the lack of alignment between the measured Doppler shifts with the radial velocities expected along the orbit of either the G2\,IV or the K0\,IV star. This is interpreted as the effect of both stars in AR\,Lac contributing to the total system X-ray emission and impacting the outcome of the net Doppler shifts as a function of the orbital phase.

\subsection{UX\,Ari}
\label{sec:uxari_data}

UX\,Ari was observed with the ACIS-S in combination with the HETG on 2000 January 12 at 12:49 (UTC) for a total exposure time of 49~ks (ObsID~605). The sourced displayed a fairly stable X-ray emission across the entire observation \citep{Huenemoerder13b}. Following the same procedure adopted for the previous sources, we first extract the $\pm$1 order MEG and HEG spectra using the entire exposure time available. We considered first the +1 MEG spectrum to determine the best fit parameters of a model comprising 4 CIE components in {\sc spex} (see Sect.~\ref{sec:uxari_intro}), including a REDS component to take into account the effect of the velocity shifts within the binary. Given the known low absorption column density in the direction of the source \citep{gudel99}, we verified that the addition of an HOT component to take into account the Galactic absorption did not significantly affect the best fit parameters within the 1~$\sigma$ associated uncertainties. 
This model gave acceptable results with no substantial residuals from the fit to all detectable lines, and the same conclusion applies when the model is applied to the remaining three spectra extracted from the same observation (MEG -1 order and HEG $\pm$1 order spectra). We also performed a fit of the MEG +1 spectrum within the {\sc xspec} environment using a model comprising four BVVAPEC components. We verified in all cases that no relevant residuals were left in correspondence of all detected emission lines. We summarize the best fit values obtained in {\sc spex} and {\sc xspec} from the MEG +1 spectrum in Table~\ref{tab:uxari_spe}. Measurements obtained from {\sc spex} and {\sc xspec} are virtually all  consistent to within the 1~$\sigma$~c.l. associated uncertainties. 
\begin{table*}
\caption{Same as Table~\ref{tab:hr1099_obs} but for the \chan/HETG observation of UX\,Ari (ObsID~605 divided here in three segments, see text for more details). The orbital phases were calculated from the ephemerides published by \citet{duemmler01}. All indicated uncertainties are at 1$\sigma$~c.l. The uncertainty on the phase includes the known uncertainties on the published source ephemerides.} 
\label{tab:uxari_obs}
\begin{tabular}{ccccccc}
\hline
 & Middle & Effective &   & Corrected velocity & Corrected velocity & Barycenter correction \\
 & Observational time & exposure & Phase & V$_{{\rm corr}_{\rm spex}}$ & V$_{{\rm corr}_{\rm xspec}}$ & V$_{\rm bary}$ \\
ObsID & (HJD)  & (ksec)   & $\phi$ & (km/s) & (km/s) & (km/s) \\
\hline
605$_{\rm int1}$  & 2451556.1378 &  16.2 & 0.007$_{-0.015}^{+0.015}$ & 72.3$_{-5.4}^{+5.4}$ & 75.2$_{-5.3}^{+5.3}$ &  24.5 \\ 
605$_{\rm int2}$  & 2451556.3385 &  16.2 & 0.038$_{-0.015}^{+0.015}$ &  59.5$_{-4.7}^{+4.7}$ & 63.2$_{-4.6}^{+4.6}$ &  24.6 \\ 
605$_{\rm int3}$ & 2451556.5392 &  16.2 & 0.069$_{-0.015}^{+0.015}$ & 54.6$_{-4.7}^{+4.7}$ & 65.1$_{-4.6}^{+4.6}$ &  24.6 \\ 
\hline
\end{tabular}
\end{table*}
\begin{table}
    \begin{center}
    \caption{\label{tab:uxari_spe} Same as Table~\ref{tab:hr1099_spe}, but for the case of the +1 order MEG spectrum of UX\,Ari obtained from the ObsID~605. The best fit was obtained by using four plasma components (CIE in {\sc spex} and BVVAPEC in {\sc xspec}). The fit in {\sc spex} also includes a REDS component.}
    \begin{tabular}{llll}
        \hline
        \hline
        Model parameter & \multicolumn{1}{c}{Best fit values} & \multicolumn{1}{c}{Best fit values} & Units \\
                        & \multicolumn{1}{c}{{\sc spex}} & \multicolumn{1}{c}{{\sc xspec}} &  \\        
        \hline
        $kT_{1}$ & 0.24$_{-0.05}^{+0.07}$ & 0.15$_{-0.03}^{+0.05}$ & keV \\
        $kT_{2}$ & 0.67$_{-0.05}^{+0.05}$ & 0.60$_{-0.05}^{+0.04}$ & keV \\
        $kT_{3}$ & 1.08$_{-0.06}^{+0.07}$ & 1.06$_{-0.05}^{+0.06}$ & keV \\
        $kT_{4}$ & 2.0$_{-0.2}^{+0.2}$ & 1.9$_{-0.1}^{+0.2}$ & keV \\
        $N_{1}$ & 1.4$_{-0.4}^{+1.4}$ & 0.02$_{-0.01}^{+0.04}$ & 10$^{59}$~m$^{-3}$  \\
        $N_{2}$ & 2.6$_{-0.5}^{+0.5}$ & 0.008$_{-0.001}^{+0.001}$ & 10$^{59}$~m$^{-3}$  \\        
        $N_{3}$ & 4.6$_{-0.7}^{+0.8}$ & 0.015$_{-0.002}^{+0.002}$ & 10$^{59}$~m$^{-3}$  \\
        $N_{4}$ & 4.6$_{-0.9}^{+0.8}$ & 0.017$_{-0.002}^{+0.002}$ & 10$^{59}$~m$^{-3}$  \\
        O & 0.9$_{-0.1}^{+0.2}$ & 0.6$_{-0.1}^{+0.1}$ & \\
        Ne & 1.5$_{-0.1}^{+0.1}$ & 1.6$_{-0.1}^{+0.1}$ & \\
        Na & 0.7$_{-0.3}^{+0.3}$ & 0.5$_{-0.3}^{+0.3}$ & \\
        Mg & 0.20$_{-0.02}^{+0.02}$ & 0.19$_{-0.02}^{+0.02}$ & \\
        Al & 0.3$_{-0.1}^{+0.1}$ & 0.3$_{-0.1}^{+0.1}$ & \\
        Si & 0.20$_{-0.02}^{+0.02}$ & 0.20$_{-0.02}^{+0.03}$ & \\
        S & 0.19$_{-0.04}^{+0.05}$ & 0.16$_{-0.02}^{+0.02}$ & \\
        Ca & 0.4$_{-0.3}^{+0.3}$ & 0.3$_{-0.3}^{+0.3}$ & \\
        Fe & 0.15$_{-0.02}^{+0.02}$ & 0.101$_{-0.008}^{+0.008}$ & \\
        Ni & 0.12$_{-0.04}^{+0.05}$ & 0.14$_{-0.05}^{+0.05}$ & \\
        $z$ & 110.2$_{-5.4}^{+5.4}$ & 112.2$_{-5.2}^{+5.3}$ & km~s$^{-1}$ \\    
        C-statistics/d.o.f. & 1196.9/1177 & 1222.7/1112 & \\
        \hline
        \hline
    \end{tabular}
\end{center}
\end{table}

\begin{figure}
  \centering
  \includegraphics[width=9.0cm]{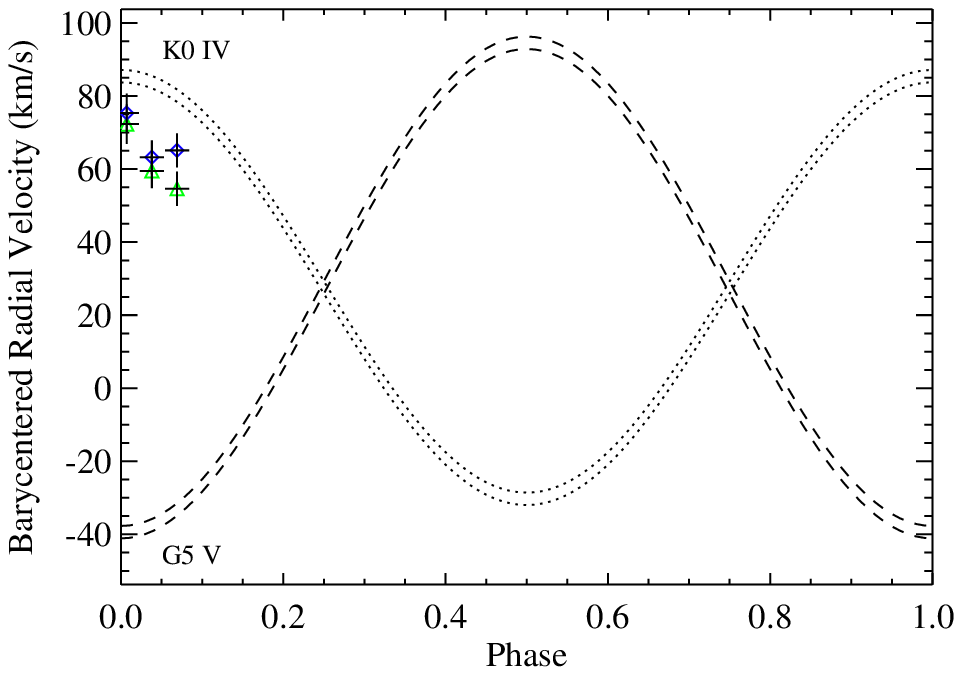}
  \caption{\label{fig:uxari_orbit} Same as Fig.~\ref{fig:hr1099_orbit} but for the case of UX\,Ari. The dotted and dashed lines are the expected barycentered radial velocities of the G2 V and K0 IV stars (see Sect.~\ref{sec:intro}) calculated according to the ephemerides published by \citet{duemmler01}. There are two dotted and two dashed lines as UX\,Ari center of mass velocity changes over time and we plotted the orbital solutions from \citet{duemmler01} using the highest and lowest reported values of this velocity. The green points correspond to values obtained from the spectral fits in {\sc spex}, while blue points were obtained from the fits in {\sc xspec}. The error bars for each radial velocity measurement are given at 1~$\sigma$~c.l. and are only statistical. The uncertainty on the phase for each measurement includes the duration of the observation as well as the uncertainties on the ephemerides indicated by \citet{duemmler01}.}  
\end{figure}

We calculated the orbital phase of the ObsID~605 by using the ephemerides published by \citet{duemmler01} and noticed that the observation lies close to the maximum velocity separation between the two system component (phase 0 in Fig.~\ref{fig:uxari_orbit}). In order to evidence a possible velocity shift change during the limited orbital phase coverage offered by the ObsID~605, we split the observation in three segments of equal exposure (about 16.2~ks) and performed a fit to the $\pm$1 MEG and HEG spectra extracted from each segment with the spectral model reported in Table~\ref{tab:uxari_spe}. As in the case of the previous sources, the velocity shift in each segment was computed as a weighted average between the values obtained from the four $\pm$1 MEG and HEG spectra. The same procedure was applied for the fits within {\sc xspec} of the same spectra. We report the outcome of this analysis in Table~\ref{tab:uxari_obs} and in Fig.~\ref{fig:uxari_orbit}. 

Interestingly, all velocity shifts measured from the {\sc spex} and {\sc xspec} fits to the spectra of the three segments lie about 20~km~s$^{-1}$ below the expected values according to the radial velocity curve of the primary K0 IV star (see Fig.~\ref{fig:uxari_orbit}). As the secondary star in UX\,Ari is also known to be chromospherically active (albeit at a level much lower than the primary), we tentatively ascribe the lower velocity shifts measured from the HETG spectra to the influence of the secondary component. In order to provide a quantitative evaluation of this possibility, we performed a spectral simulation where both the primary and secondary components are assumed to have the same spectral energy distribution and participate with different levels in the overall X-ray emission. Although this is a simplified approach because at present it is not possible to disentangle from the \chan\ data the contribution of the two stars, our simulation revealed that a contribution to the total emission from the secondary component by as little as 10\% is already able to produce the decrease in the velocity shifts reported in Fig.~\ref{fig:uxari_orbit}. Such a modest contribution seems roughly in line with the literature reports about the presence of a limited chromospherical activity from the G5 V star (see Sect.~\ref{sec:uxari_intro}).

\subsection{V824\,Ara}
\label{sec:v824ara_data}

V824\,Ara was observed with the ACIS-S in combination with the HETG on 2002 July 10 at 11:27 (UTC) for a total exposure time of 94~ks (ObsID~605). 
The sourced displayed a noticeable X-ray variability during the observation, with a flare-like structure extending in the time interval 25~ks-45~ks since the beginning of the exposure \citep{lalitha15}. Following the same procedure adopted for the previous sources, we first extract the $\pm$1 order MEG and HEG spectra using the entire exposure time available. We forced the {\sc chandra\_repro} script to use a customized extraction region to avoid the contamination of the spectra from the inner system binary from those of the close-by M3 star (see Sect.~\ref{sec:v824ara_intro}) by setting the parameter {\sc width\_factor\_hetg=8}. We considered first the +1 HEG spectrum to determine the best fit parameters of a model comprising 3 CIE components in {\sc spex} (see Sect.~\ref{sec:v824ara_intro}), including a REDS component to take into account the effect of the velocity shifts within the binary. As we could not find published data about the absorption column density in the direction to the source measured via X-ray spectroscopy, we added to the fit a HOT component with an upper limit on the column density estimated from the on-line {\sc heasarc} tool of 5.4$\times$10$^{20}$~cm$^{-2}$ (see Sect.~\ref{sec:impeg_data}). 
This model also gave acceptable results for the fits to the other HEG and MEG $\pm$1 order spectra of the same observation, and we report the best fit parameters in Table~\ref{tab:v824ara_spe} (see also Fig.~\ref{fig:v824_plot_spex} for a graphical representation of the source spectrum, the best fit, and the residuals from the fit). We checked that the adopted absorption column density did not significantly impact the parameters for the 3 CIE components derived from the fits. As for all other sources, we also performed a fit of the HEG +1 spectrum within the {\sc xspec} environment using a model comprising three BVVAPEC components (see Table~\ref{tab:v824ara_spe} and Fig.~\ref{fig:v824_plot_spex}). Measurements obtained from {\sc spex} and {\sc xspec} are all  consistent to within the 1~$\sigma$~c.l. associated uncertainties.
\begin{table*}
\caption{Same as Table~\ref{tab:hr1099_obs} but for the \chan/HETG observation of V824\,Ara (ObsID~2538 divided here in six  segments, see text for more details). The orbital phases were calculated from the ephemerides published by \citet{strassmeier00b}. All indicated uncertainties are at 1$\sigma$~c.l. The uncertainty on the phase includes the known uncertainties on the published source ephemerides.} 
\label{tab:v824ara_obs}
\begin{tabular}{ccccccc}
\hline
 & Middle & Effective &   & Corrected velocity & Corrected velocity & Barycenter correction \\
 & Observational time & exposure & Phase & V$_{{\rm corr}_{\rm spex}}$ & V$_{{\rm corr}_{\rm xspec}}$ & V$_{\rm bary}$ \\
ObsID & (HJD)  & (ksec)   & $\phi$ & (km/s) & (km/s) & (km/s) \\
\hline
2538$_{\rm int1}$  & 2452466.0824 &  15.7 & 0.383$_{-0.055}^{+0.055}$ & -15.1$_{-5.5}^{+5.5}$ & -12.7$_{-5.2}^{+5.2}$ &  9.0 \\ 
2538$_{\rm int2}$  & 2452466.2665 &  15.7 & 0.492$_{-0.055}^{+0.055}$ &  -11.7$_{-5.9}^{+5.9}$ & -12.3$_{-5.6}^{+5.6}$ &  9.0 \\ 
2538$_{\rm int3}$ & 2452466.4506 &  15.7 & 0.602$_{-0.055}^{+0.055}$ & 0.9$_{-5.6}^{+5.6}$ & 2.8$_{-5.4}^{+5.4}$ &  9.1 \\ 
2538$_{\rm int4}$ & 2452466.6346 &  15.7 & 0.711$_{-0.055}^{+0.055}$ & 0.5$_{-5.3}^{+5.3}$ & 0.1$_{-5.4}^{+5.4}$ &  9.1 \\ 
2538$_{\rm int5}$ & 2452466.8187 &  15.7 & 0.821$_{-0.055}^{+0.055}$ & 11.8$_{-5.4}^{+5.4}$ & 14.0$_{-4.8}^{+4.8}$ &  9.2 \\ 
2538$_{\rm int6}$ & 2452467.0028 &  15.7 & 0.930$_{-0.055}^{+0.055}$ & 26.6$_{-5.1}^{+5.1}$ & 21.5$_{-4.6}^{+4.6}$ &  9.3 \\ 
\hline
\end{tabular}
\end{table*}
\begin{table}
    \begin{center}
    \caption{Same as Table~\ref{tab:hr1099_spe}, but for the case of the +1 order HEG spectrum of V824\,Ara obtained from the ObsID~2538. The best fit was obtained by using three plasma components (CIE in {\sc spex} and BVVAPEC in {\sc xspec}). The fit in {\sc spex} also includes a REDS component.}
    \begin{tabular}{llll}
        \hline
        \hline
        Model parameter & \multicolumn{1}{c}{Best fit values} & \multicolumn{1}{c}{Best fit values} & Units \\
                        & \multicolumn{1}{c}{{\sc spex}} & \multicolumn{1}{c}{{\sc xspec}} &  \\        
        \hline
        $kT_{1}$ & 0.41$_{-0.02}^{+0.02}$ & 0.44$_{-0.02}^{+0.02}$ & keV \\
        $kT_{2}$ & 0.88$_{-0.03}^{+0.03}$ & 0.91$_{-0.02}^{+0.02}$ & keV \\
        $kT_{3}$ & 1.83$_{-0.09}^{+0.09}$ & 1.86$_{-0.08}^{+0.08}$ & keV \\
        $N_{1}$ & 1.1$_{-0.1}^{+0.1}$ & 0.009$_{-0.001}^{+0.002}$ & 10$^{59}$~m$^{-3}$  \\
        $N_{2}$ & 1.4$_{-0.1}^{+0.1}$ & 0.012$_{-0.001}^{+0.001}$ & 10$^{59}$~m$^{-3}$  \\        
        $N_{3}$ & 1.4$_{-0.1}^{+0.1}$ & 0.012$_{-0.001}^{+0.001}$ & 10$^{59}$~m$^{-3}$  \\
        O & 0.6$_{-0.2}^{+0.2}$ & 0.27$_{-0.08}^{+0.10}$ & \\
        Ne & 0.80$_{-0.08}^{+0.08}$ & 0.85$_{-0.08}^{+0.09}$ & \\
        Na & 0.3$_{-0.3}^{+0.4}$ & 0.12$_{-0.12}^{+0.33}$ & \\
        Mg & 0.36$_{-0.03}^{+0.04}$ & 0.33$_{-0.03}^{+0.03}$ & \\
        Al & 0.8$_{-0.2}^{+0.2}$ & 0.7$_{-0.1}^{+0.1}$ & \\
        Si & 0.31$_{-0.02}^{+0.02}$ & 0.30$_{-0.03}^{+0.03}$ & \\
        S & 0.27$_{-0.06}^{+0.06}$ & 0.23$_{-0.05}^{+0.06}$ & \\
        Ca & 0.8$_{-0.3}^{+0.3}$ & 0.7$_{-0.3}^{+0.3}$ & \\
        Fe & 0.36$_{-0.03}^{+0.03}$ & 0.21$_{-0.02}^{+0.02}$ & \\
        Ni & 0.46$_{-0.08}^{+0.09}$ & 0.37$_{-0.07}^{+0.08}$ & \\
        $z$ & 22.0$_{-4.8}^{+4.8}$ & 18.9$_{-4.8}^{+4.8}$ & km~s$^{-1}$ \\    
        C-statistics/d.o.f. & 2186.0/1739 & 2246.4/1783 & \\
        \hline
        \hline
    \end{tabular}\label{tab:v824ara_spe}
\end{center}
\end{table}

\begin{figure}
  \centering
  \includegraphics[width=9.0cm]{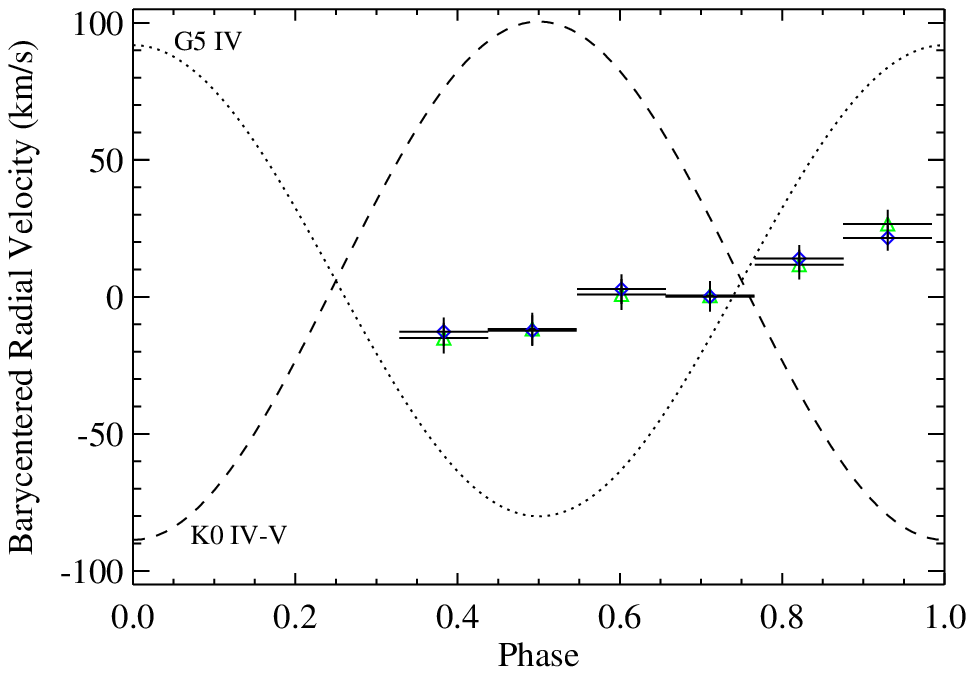}
  \caption{\label{fig:v824ara_orbit} Same as Fig.~\ref{fig:hr1099_orbit} but for the case of V824\,Ara. The dotted and dashed lines are the expected barycentered radial velocities of the primary (G5 IV) and secondary (K0 IV-V) stars calculated according to the ephemerides published by \citet{strassmeier00b}. The green points correspond to radial velocity values obtained from the spectral fits in {\sc spex}, while blue points were obtained from the fits in {\sc xspec}. The error bars for each radial velocity measurement are given at 1~$\sigma$~c.l. and are only statistical. The uncertainty on the phase for each measurement includes the duration of the observation as well as the uncertainties on the ephemerides indicated by \citet{strassmeier00b}.}  
\end{figure}

In order to search for possible velocity shift changes during the orbital phase coverage offered by the ObsID~2538, we split the observation in six segments of equal exposure (about 15.7~ks) and performed a fit to the $\pm$1 MEG and HEG spectra extracted from each segment with the spectral model reported in Table~\ref{tab:v824ara_spe}. As in the case of the previous sources, the velocity shift in each segment was computed as a weighted average between the values obtained from the four $\pm$1 MEG and HEG spectra. We report the outcome of this analysis in Table~\ref{tab:v824ara_obs} and in Fig.~\ref{fig:v824ara_orbit}. 

Looking at Fig.~\ref{fig:v824ara_orbit}, there seems to be a lack of clear alignment of the radial velocities along either the values expected for the G5~IV star or the K0~IV-V star. As in a few other systems analyzed in this work, the most likely reason is that also in V824\,Ara both stars are active in X-rays. The measured shifts suggest that G5\,IV is possibly slightly dominating the the overall emission.

\subsection{TZ\,CrB}
\label{sec:tzcrb_data}

TZ\,CrB was observed by \chan\ using the ACIS-S in combination with the HETG on 2000 June 18 at 13:41 (UTC) for a total exposure time of 84.8~ks. This observation (ObsID~15) provides a coverage of about 86\% of the total binary orbit and it is the same dataset that was previously analyzed by \citet{osten03} using a line-based technique (see Sect.~\ref{sec:tzcrb_intro}). The sourced displayed a bright flare toward the end of the observation, which decay was only partly covered by the \chan\ data \citep{Huenemoerder13b, osten03}. We first noticed from the ACIS-S image that a relatively faint source is visible close to the main target and separated by roughly $\sim$7\arcsec.\ The position of this source matches that of the known $\sigma^1$\,CrB, confirming the previous private communication from \citet{osten03} mentioned in the paper by \citet{suh05}. As we could not find any detail about the contamination of such object onto the TZ\,CrB HETG spectra in \citet{osten03}, we performed here an independent evaluation. We extracted the ACIS-S zero order spectra of both TZ\,CrB and $\sigma^1$\,CrB, together with the corresponding response and ancillary files, using the standard procedure and the {\sc CIAO} tool {\sc specextract}\footnote{See \url{https://cxc.cfa.harvard.edu/ciao/ahelp/specextract.html}.}. Loading these spectra in {\sc xspec}, we obtained a net count-rate of 0.158$\pm$0.001~cts~s$^{-1}$ for TZ\,CrB and 0.0098$\pm$0.0003~cts~s$^{-1}$ for $\sigma^1$\,CrB. The two spectra are also remarkably  different, with the emission from $\sigma^1$\,CrB being much softer than that of TZ\,CrB and falling below the instrument background level at energies $\gtrsim$2.0~keV. As the two sources are only separated by a few arcseconds, it is impossible to disentangle the contribution of $\sigma^1$\,CrB in the dispersed HETG spectra of TZ\,CrB. In the following analysis, we thus have to take into account that results we report below on TZ\,CrB are inevitably affected by the contribution from $\sigma^1$\,CrB at a level of roughly 6\%. The contamination occurs virtually only below 2~keV, but this is also the energy band where the bulk of the emission lines from TZ\,CrB used for the determination of the velocity shifts in the binary are recorded. Having clarified the contamination, we extracted the $\pm$1 order MEG and HEG spectra using the entire exposure time available during the ObsID~15. We considered first the +1 MEG spectrum to determine the best fit parameters of a model comprising four CIE components in {\sc spex} (see Sect.~\ref{sec:tzcrb_intro}). We also included a REDS component and a HOT component to take into account the velocity shifts in the binary and the interstellar absorption, respectively \citep[the value of the absorption column density in the direction of the source was fixed at 2.5$\times$10$^{18}$~cm$^{-2}$ and checked to have no effect on the spectral results; see][]{forcada03}. This model gave acceptable results also for the fits to the other HEG and MEG $\pm$1 order spectra of the same observation, and we report the best fit parameters in Table~\ref{tab:tzcrb_spe}. We also performed a fit of the MEG +1 spectrum within the {\sc xspec} environment using a model comprising four BVVAPEC components (see Table~\ref{tab:tzcrb_spe}). A visual representation of the source MEG +1  spectrum fit with the above best model within both the {\sc spex} and {\sc xspec} environments, including the residuals from these fits, is provided in Fig.~\ref{fig:tzcrb_plot_spex}.
\begin{table*}
\caption{Same as Table~\ref{tab:hr1099_obs} but for the \chan/HETG observation of TZ\,CrB (ObsID~15 divided here in five segments, see text for more details). The orbital phases were calculated from the ephemerides published by \citet{deepak09}. All indicated uncertainties are at 1$\sigma$~c.l. The uncertainty on the phase includes the known uncertainties on the published source ephemerides.} 
\label{tab:tzcrb_obs}
\begin{tabular}{ccccccc}
\hline
 & Middle & Effective &   & Corrected velocity & Corrected velocity & Barycenter correction \\
 & Observational time & exposure & Phase & V$_{{\rm corr}_{\rm spex}}$ & V$_{{\rm corr}_{\rm xspec}}$ & V$_{\rm bary}$ \\
ObsID & (HJD)  & (ksec)   & $\phi$ & (km/s) & (km/s) & (km/s) \\
\hline
15$_{\rm int1}$  & 2451714.18629 &  16.7 & 0.981$_{-0.085}^{+0.085}$ & -13.1$_{-3.2}^{+3.2}$ & -10.7$_{-3.0}^{+3.0}$ &  10.1 \\ 
15$_{\rm int2}$  & 2451714.38258 &  16.7 & 0.153$_{-0.085}^{+0.085}$ &  -19.3$_{-3.0}^{+3.0}$ & -17.5$_{-3.1}^{+3.1}$ &  10.1 \\ 
15$_{\rm int3}$ & 2451714.57888 &  16.7 & 0.325$_{-0.085}^{+0.085}$ & -9.4$_{-3.1}^{+3.1}$ & -6.5$_{-2.9}^{+2.9}$ &  10.2 \\ 
15$_{\rm int4}$ & 2451714.77517 &  16.7 & 0.498$_{-0.085}^{+0.085}$ & -9.4$_{-3.0}^{+3.0}$ & -11.5$_{-3.0}^{+3.0}$ &  10.2 \\ 
15$_{\rm int5}$ & 2451714.97146 &  16.7 & 0.670$_{-0.085}^{+0.085}$ & -6.3$_{-2.8}^{+2.8}$ & -11.2$_{-2.7}^{+2.7}$ &  10.2 \\ 
\hline
\end{tabular}
\end{table*}
\begin{table}
    \begin{center}
    \caption{Same as Table~\ref{tab:hr1099_spe}, but for the case of the +1 order MEG spectrum of TZ\,CrB obtained from the ObsID~15. The best fit was obtained by using four plasma components (CIE in {\sc spex} and BVVAPEC in {\sc xspec}). The fit in {\sc spex} also includes a REDS component.}
    \begin{tabular}{llll}
        \hline
        \hline
        Model parameter & \multicolumn{1}{c}{Best fit values} & \multicolumn{1}{c}{Best fit values} & Units \\
                        & \multicolumn{1}{c}{{\sc spex}} & \multicolumn{1}{c}{{\sc xspec}} &  \\        
        \hline
        $kT_{1}$ & 0.134$_{-0.005}^{+0.005}$ & 0.117$_{-0.007}^{+0.007}$ & keV \\
        $kT_{2}$ & 0.54$_{-0.01}^{+0.01}$ & 0.55$_{-0.01}^{+0.01}$ & keV \\
        $kT_{3}$ & 0.99$_{-0.01}^{+0.01}$ & 1.00$_{-0.01}^{+0.01}$ & keV \\
        $kT_{4}$ & 2.36$_{-0.08}^{+0.08}$ & 2.30$_{-0.08}^{+0.08}$ & keV \\        
        $N_{1}$ & 7.8$_{-1.4}^{+1.8}$ & 0.3$_{-0.1}^{+0.1}$ & 10$^{59}$~m$^{-3}$  \\
        $N_{2}$ & 0.94$_{-0.05}^{+0.05}$ & 0.0185$_{-0.0009}^{+0.0010}$ & 10$^{59}$~m$^{-3}$  \\        
        $N_{3}$ & 1.24$_{-0.05}^{+0.05}$ & 0.0237$_{-0.0009}^{+0.0009}$ & 10$^{59}$~m$^{-3}$  \\
        $N_{4}$ & 1.19$_{-0.04}^{+0.04}$ & 0.0210$_{-0.0007}^{+0.0008}$ & 10$^{59}$~m$^{-3}$  \\        
        O & 0.73$_{-0.08}^{+0.09}$ & 0.49$_{-0.06}^{+0.06}$ & \\
        Ne & 0.86$_{-0.04}^{+0.04}$ & 0.87$_{-0.04}^{+0.04}$ & \\
        Na & 0.8$_{-0.2}^{+0.2}$ & 0.6$_{-0.2}^{+0.2}$ & \\
        Mg & 0.56$_{-0.02}^{+0.03}$ & 0.49$_{-0.02}^{+0.02}$ & \\
        Al & 0.65$_{-0.08}^{+0.08}$ & 0.53$_{-0.07}^{+0.08}$ & \\
        Si & 0.42$_{-0.02}^{+0.02}$ & 0.38$_{-0.02}^{+0.02}$ & \\
        S & 0.36$_{-0.04}^{+0.04}$ & 0.33$_{-0.03}^{+0.03}$ & \\
        Ca & 1.1$_{-0.3}^{+0.3}$ & 1.0$_{-0.2}^{+0.3}$ & \\
        Fe & 0.47$_{-0.02}^{+0.02}$ & 0.29$_{-0.01}^{+0.01}$ & \\
        Ni & 0.49$_{-0.04}^{+0.04}$ & 0.41$_{-0.04}^{+0.04}$ & \\
        $z$ & 8.1$_{-2.8}^{+2.8}$ & 7.2$_{-2.7}^{+2.7}$ & km~s$^{-1}$ \\    
        C-statistics/d.o.f. & 2384.1/1357 & 2756.0/1289 & \\
        \hline
        \hline
    \end{tabular}\label{tab:tzcrb_spe}
\end{center}
\end{table}

\begin{figure}
  \centering
  \includegraphics[width=9.0cm]{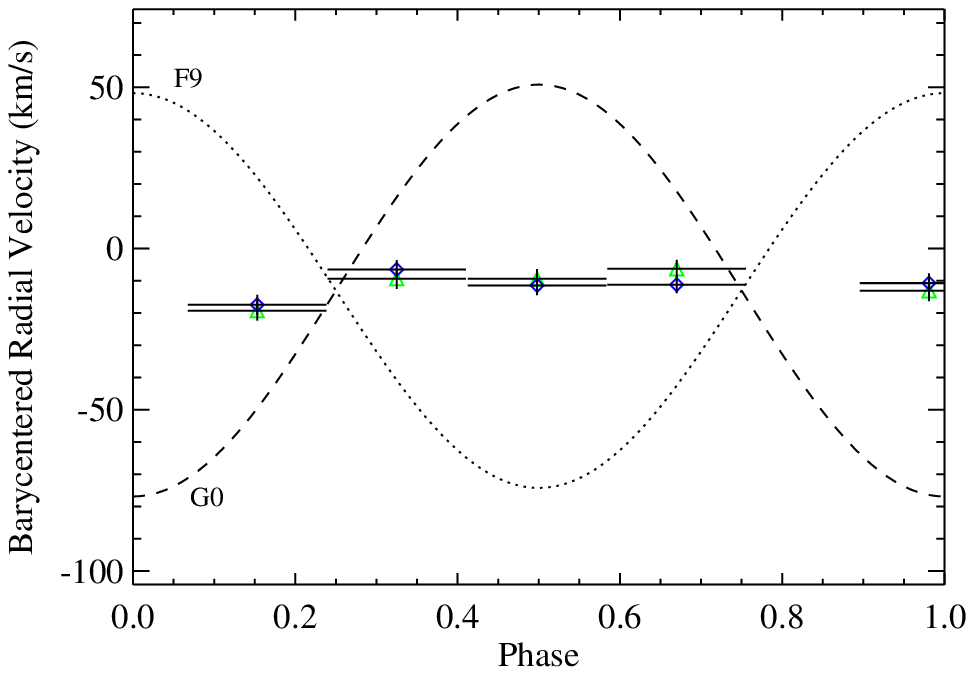}
  \caption{\label{fig:tzcrb_orbit} Same as Fig.~\ref{fig:hr1099_orbit} but for the case of TZ\,CrB. The dotted and dashed lines are the expected barycentered radial velocities of the primary (F9-G0) and secondary (G0-G1) stars calculated according to the ephemerides published by \citet{deepak09}. The green points correspond to radial velocity values obtained from the spectral fits in {\sc spex}, while blue points were obtained from the fits in {\sc xspec}. The error bars for each radial velocity measurement are given at 1~$\sigma$~c.l. and are only statistical. The uncertainty on the phase for each measurement includes the duration of the observation as well as the uncertainties on the ephemerides indicated by \citet{deepak09}.}  
\end{figure}

In order to search for possible velocity shift changes during the orbital phase coverage offered by the ObsID~15, we split the observation in five segments of equal exposure (about 16.7~ks) and performed a fit to the $\pm$1 MEG and HEG spectra extracted from each segment with the spectral model reported in Table~\ref{tab:tzcrb_spe}. The velocity shift in each segment was computed as a weighted average between the values obtained from the four $\pm$1 MEG and HEG spectra. We report the outcome of this analysis in Table~\ref{tab:tzcrb_obs} and in Fig.~\ref{fig:tzcrb_orbit}. 

Looking at Fig.~\ref{fig:tzcrb_orbit}, it is evident that velocity shift measurements obtained from {\sc spex} and {\sc xspec} are all  consistent to within the 1~$\sigma$~c.l. associated uncertainties. Although the results obtained for TZ\,CrB could be marginally affected by the contamination from the nearby source $\sigma^1$\,CrB, it seems rather clear that the measured velocities in both fitting environments do not show an evident correlation with either the expected velocity curve of the primary or secondary star. This is compatible with the findings in the literature which reported evidences that both stellar components in TZ\,CrB are chromospherically active \citep{frasca97, osten03}.

\subsection{HR\,5110}
\label{sec:hr5110_data}

HR\,5110 has been observed by the ACIS-S in combination with the HETG twice, the first time on 2008 April 4th at 01:34 (UTC; ObsID~8934) and the second time on 2008 April 06 at 02:47 (UTC; ObsID~8935). The exposure times of the two observations were of 38.9~ks and 40.0~ks, respectively. The sourced displayed a fairly constant X-ray luminosity across the exposures. Both observations have not yet been reported elsewhere in the literature. Following the same procedure of all other sources, we processed the data with the {\sc chandra\_repro} script and extracted first the average MEG +1 spectrum of the source in the ObsID~8934. We found that a model comprising 3 plasma components in both {\sc spex} and {\sc xspec} could fit the data reasonably well, with no evident line left untreated in the residuals (the usual REDS component was added to the model used within {\sc spex} in order to take into account velocity shifts within the binary; see Fig.~\ref{fig:hr5110_plot_spex}). We added to the fits an absorption component with a column density of 9$\times$10$^{19}$~cm$^{-2}$ (see Sect.~\ref{sec:hr5110_intro}) and verified that such component does not significantly affect the parameters of the plasma components. The same model used to describe the MEG +1 spectrum could satisfactorily fit also the MEG -1 spectrum, as well as the HEG $\pm$1 spectra. The same model was also applied to the ObsID~8935, obtaining equivalently acceptable results. We report the results of this analysis in Table~\ref{tab:hr5110_spe}. 

\begin{figure}
  \centering
  \includegraphics[width=9.0cm]{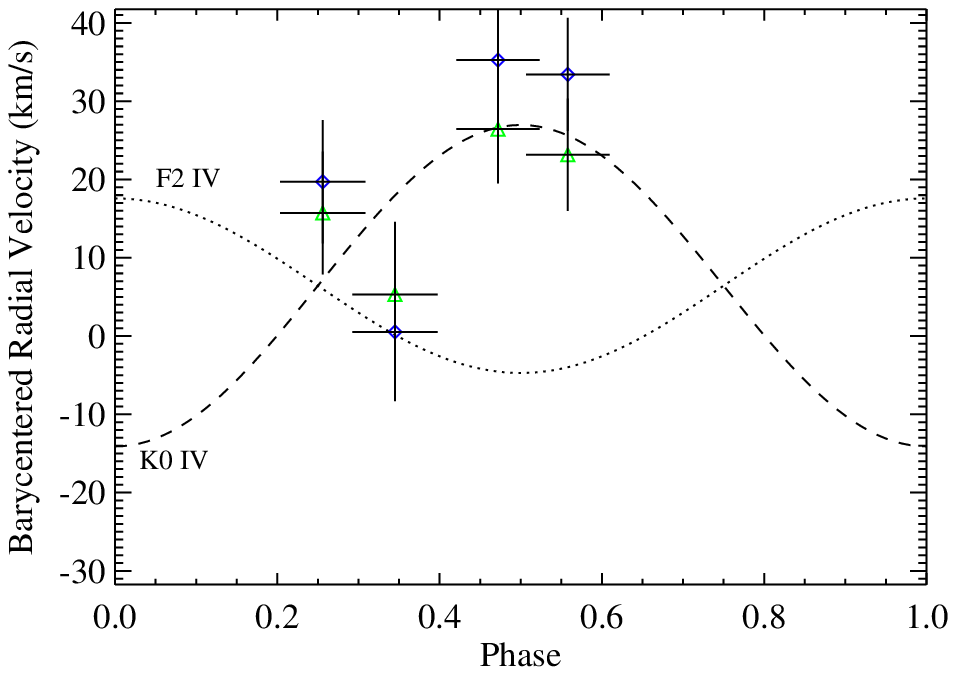}
  \caption{\label{fig:hr5110_orbit} Same as Fig.~\ref{fig:hr1099_orbit} but for the case of HR\,5110. The dotted and dashed lines are the expected barycentered radial velocities of the primary (F2\,IV) and secondary (K0\,IV) stars calculated according to the ephemerides published by \citet{eker87} and \citet{ransom03}. The green points correspond to radial velocity values obtained from the spectral fits in {\sc spex}, while blue points were obtained from the fits in {\sc xspec}. The error bars for each radial velocity measurement are given at 1~$\sigma$~c.l. and are only statistical. The uncertainty on the phase for each measurement includes the duration of the observation as well as the uncertainties on the ephemerides indicated by \citet{eker87} and \citet{ransom03}.}  
\end{figure}

In order to investigate possible velocity shifts within the system, we split both ObsID~8934 and 8935 in two segments and fit all spectra with the same model described above, leaving in each case only the velocity shift and the normalizations of the three plasma components to vary (as done in the previous sections). The same fits were carried out in both the {\sc spex} and {\sc xspec} environments, deriving for each segment in each environment the velocity shift measurement as a weighted average of the shift in each of the four MEG and HEG $\pm$1 spectra. The results of this analysis is summarized in Fig.~\ref{fig:hr5110_orbit} and in Table~\ref{tab:hr5110_obs}. The latter table also reports the computed orbital phases of each segment by using the ephemerides published by \citet{eker87}. We also made use of the latest update on the system orbital period in \citet{ransom03}. As it can be seen from Fig.~\ref{fig:hr5110_orbit}, the obtained velocity shifts seem in good agreement with those expected along the orbit of the secondary star, although the measurement around phases 0.3-0.4 is somewhat lower than expected. Measurements obtained from {\sc spex} and {\sc xspec} are all  consistent to within the $\sim$1~$\sigma$~c.l. associated uncertainties. 
\begin{table*}
\caption{Same as Table~\ref{tab:hr1099_obs} but for the \chan/HETG observation of HR\,5110 (ObsID~8934 and 8935, each divided in two segments, see text for more details). The orbital phases were calculated from the ephemerides published by \citet{eker87} with the most updated orbital period from \citet{ransom03}. All indicated uncertainties are at 1$\sigma$~c.l. The uncertainty on the phase includes the known uncertainties on the published source ephemerides.} 
\label{tab:hr5110_obs}
\begin{tabular}{ccccccc}
\hline
 & Middle & Effective &   & Corrected velocity & Corrected velocity & Barycenter correction \\
 & Observational time & exposure & Phase & V$_{{\rm corr}_{\rm spex}}$ & V$_{{\rm corr}_{\rm xspec}}$ & V$_{\rm bary}$ \\
ObsID & (HJD)  & (ksec)   & $\phi$ & (km/s) & (km/s) & (km/s) \\
\hline
8934$_{\rm int1}$  & 2454560.7010 &  19.2 & 0.222$_{-0.051}^{+0.051}$ & 35.3$_{-7.1}^{+7.1}$ & 26.4$_{-7.0}^{+7.0}$ &  3.6 \\ 
8934$_{\rm int1}$  & 2454560.9263 &  19.2 & 0.308$_{-0.051}^{+0.051}$ & 33.4$_{-7.2}^{+7.2}$ & 23.2$_{-7.2}^{+7.2}$ &  3.7 \\ 
8935$_{\rm int2}$  & 2454562.7493 &  19.8 & 0.006$_{-0.053}^{+0.053}$ & 19.7$_{-7.9}^{+7.9}$ & 15.7$_{-7.8}^{+7.8}$ &  4.4 \\ 
8935$_{\rm int2}$  & 2454562.9811 &  19.8 & 0.095$_{-0.053}^{+0.053}$ & 0.5$_{-8.9}^{+8.9}$ & 5.3$_{-9.3}^{+9.3}$ &  4.4 \\ 
\hline
\end{tabular}
\end{table*}
\begin{table}
    \begin{center}
    \caption{Same as Table~\ref{tab:hr1099_spe}, but for the case of the +1 order MEG spectrum of HR\,5110 obtained from the ObsID~8934. The best fit was obtained by using three plasma components (CIE in {\sc spex} and BVVAPEC in {\sc xspec}). The fit in {\sc spex} also includes a REDS component.}
    \begin{tabular}{llll}
        \hline
        \hline
        Model parameter & \multicolumn{1}{c}{Best fit values} & \multicolumn{1}{c}{Best fit values} & Units \\
                        & \multicolumn{1}{c}{{\sc spex}} & \multicolumn{1}{c}{{\sc xspec}} &  \\        
        \hline
        $kT_{1}$ & 0.42$_{-0.03}^{+0.03}$ & 0.44$_{-0.03}^{+0.03}$ & keV \\
        $kT_{2}$ & 0.97$_{-0.05}^{+0.05}$ & 0.95$_{-0.04}^{+0.06}$ & keV \\
        $kT_{3}$ & 2.0$_{-0.2}^{+0.2}$ & 1.9$_{-0.1}^{+0.1}$ & keV \\
        $N_{1}$ & 1.6$_{-0.3}^{+0.3}$ & 0.005$_{-0.001}^{+0.001}$ & 10$^{59}$~m$^{-3}$  \\
        $N_{2}$ & 2.2$_{-0.3}^{+0.3}$ & 0.0079$_{-0.0009}^{+0.0009}$ & 10$^{59}$~m$^{-3}$  \\        
        $N_{3}$ & 2.8$_{-0.4}^{+0.4}$ & 0.015$_{-0.001}^{+0.001}$ & 10$^{60}$~m$^{-3}$  \\
        O & 0.5$_{-0.2}^{+0.2}$ & 0.3$_{-0.1}^{+0.1}$ & \\
        Ne & 0.68$_{-0.08}^{+0.009}$ & 0.8$_{-0.1}^{+0.10}$ & \\
        Mg & 0.32$_{-0.04}^{+0.04}$ & 0.28$_{-0.04}^{+0.04}$ & \\
        Al & 0.6$_{-0.2}^{+0.2}$ & 0.6$_{-0.2}^{+0.2}$ & \\
        Si & 0.27$_{-0.03}^{+0.03}$ & 0.25$_{-0.03}^{+0.04}$ & \\
        S & 0.28$_{-0.08}^{+0.09}$ & 0.26$_{-0.07}^{+0.08}$ & \\
        Ca & 1.1$_{-0.6}^{+0.7}$ & 1.3$_{-0.6}^{+0.7}$ & \\
        Fe & 0.24$_{-0.03}^{+0.03}$ & 0.16$_{-0.02}^{+0.02}$ & \\
        Ni & 0.24$_{-0.09}^{+0.10}$ & 0.13$_{-0.08}^{+0.10}$ & \\
        $z$ & -3.4$_{-12.6}^{+12.6}$ & 4.6$_{-15.4}^{+15.4}$ & km~s$^{-1}$ \\    
        C-statistics/d.o.f. & 1154.5/1034 & 1023.8/920 & \\
        \hline
        \hline
    \end{tabular}\label{tab:hr5110_spe}
\end{center}
\end{table}

\subsection{$\sigma$\,Gem}
\label{sec:sigmagem_data}

\chan\ observed with the ACIS-S in combination with the HETG the source $\sigma$\,Gem in two occasions. The first was during the ObsID~5422, carried out on 2005 May 16 at 11:57 (UTC) for a total exposure time of 63.9~ks, while the second was during the ObsID~6282 carried out on 2005 May 17 at 17:24 (UTC) for a total exposure time of 58.9~ks. During both exposures, the source was found to be slowly decaying from a flare \citep{huenemoerder13}. We analyzed first the MEG +1 spectrum from the ObsID~5422 and found that it could be well described by using a model comprising 3 plasma components (see Fig.~\ref{fig:sigmagem_plot_spex}). We performed as in all other cases the fits in both the {\sc spex} and {\sc xspec} environments. Within {\sc spex}, we included in the model 3 CIE components for the plasma, a REDS component to take into account the velocity shifts within the binary, and a HOT component to mimic the effect of the Galactic absorption (we fixed the column density value at 2$\times$10$^{22}$~cm$^{-2}$, see Sect.~\ref{sec:sigmagem_intro}). The same model was used to fit also the data of the remaining three spectra in ObsID~5422 (MEG -1 and HEG $\pm$1), as well as the spectra from the ObsID~6282. We only left in these fits the normalizations of the CIE components and the velocity shift values free to vary. In all cases, we obtained fully acceptable results with no relevant residuals around all significant emission lines. We also tested that in all cases leaving the temperatures of the CIE components free to vary in the fits did not provide any significant improvement and also did not affect the measured values of the velocity shifts (to within the associated uncertainties). The same procedure was applied to the fits within {\sc xspec}. For each of the two HETG observations, we determined the value of the velocity shift as the weighted average of the four measurements obtained from the MEG $\pm$1 and HEG $\pm$1 spectra. 
 \begin{figure}
  \centering
  \includegraphics[width=9.0cm]{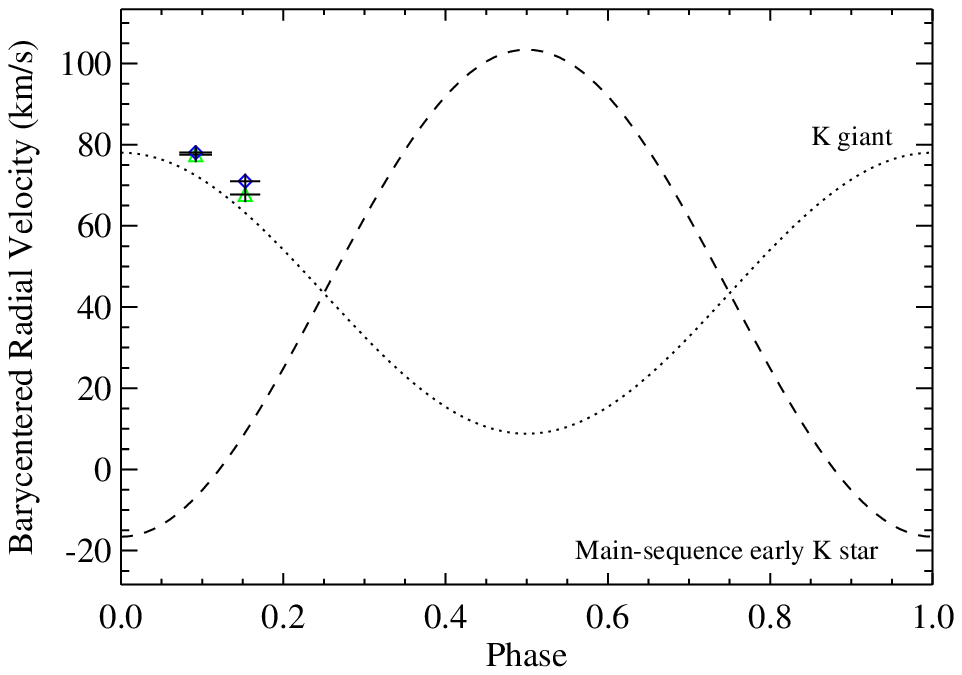}
  \caption{\label{fig:sigmagem_orbit} Same as Fig.~\ref{fig:hr1099_orbit} but for the case of $\sigma$\,Gem. The dotted and dashed lines are the expected barycentered radial velocities of the primary (K giant) and secondary (main-sequence early K star) stars calculated according to the ephemerides published by \citet{rotten15}. The green points correspond to radial velocity values obtained from the spectral fits in {\sc spex}, while blue points were obtained from the fits in {\sc xspec}. The error bars for each radial velocity measurement are given at 1~$\sigma$~c.l. and are only statistical. The uncertainty on the phase for each measurement includes the duration of the observation as well as the uncertainties on the ephemerides indicated by \citet{rotten15}.}  
\end{figure}

We report in Table~\ref{tab:sigmagem_spe} the best fit parameters obtained for the MEG +1 spectrum in ObsID~5422 obtained from fits within both {\sc spex} and {\sc xspec}. Table~\ref{tab:sigmagem_obs} and Fig.~\ref{fig:sigmagem_orbit} show the details of the HETG observations, including the phase determination obtained using the ephemerides from \citet{rotten15} and the velocity shifts obtained from the {\sc spex} and {\sc xspec} fits. Measurements obtained from {\sc spex} and {\sc xspec} are all  consistent to within the 1~$\sigma$~c.l. associated uncertainties. Looking at Fig.~\ref{fig:sigmagem_orbit}, we conclude that the primary star in the system is by far the main contributor to the X-ray emission, as the HETG velocity shifts match quite accurately the predictions along the orbit of the K giant. Given the relatively long orbital period of the system, about 19.6~days, the two HETG observations provided only a limited coverage in orbital phase ($\leq$0.02) and thus we did not split these observations in sub-segments. 
\begin{table*}
\caption{Same as Table~\ref{tab:hr1099_obs} but for the \chan/HETG observations of $\sigma$\,Gem (ObsID~5422 and 6282). The orbital phases were calculated from the ephemerides published by \citet{rotten15}. All indicated uncertainties are at 1$\sigma$~c.l. The uncertainty on the phase includes the known uncertainties on the published source ephemerides.} 
\label{tab:sigmagem_obs}
\begin{tabular}{ccccccc}
\hline
 & Middle & Effective &   & Corrected velocity & Corrected velocity & Barycenter correction \\
 & Observational time & exposure & Phase & V$_{{\rm corr}_{\rm spex}}$ & V$_{{\rm corr}_{\rm xspec}}$ & V$_{\rm bary}$ \\
ObsID & (HJD)  & (ksec)   & $\phi$ & (km/s) & (km/s) & (km/s) \\
\hline
5422  & 2453507.3763 &  62.8 & 0.092$_{-0.020}^{+0.020}$ & 77.5$_{-1.8}^{+1.8}$ & 78.0$_{-1.8}^{+1.8}$ &  24.6 \\ 
6282  & 2453508.5748 &  57.9 & 0.153$_{-0.019}^{+0.019}$ & 67.7$_{-1.9}^{+1.9}$ & 71.0$_{-1.9}^{+1.9}$ &  24.2 \\ 
\hline
\end{tabular}
\end{table*}
\begin{table}
    \begin{center}
    \caption{Same as Table~\ref{tab:hr1099_spe}, but for the case of the +1 order MEG spectrum of $\sigma$\,Gem obtained from the ObsID~5422. The best fit was obtained by using three plasma components (CIE in {\sc spex} and BVVAPEC in {\sc xspec}). The fit in {\sc spex} also includes a REDS component.}
    \begin{tabular}{llll}
        \hline
        \hline
        Model parameter & \multicolumn{1}{c}{Best fit values} & \multicolumn{1}{c}{Best fit values} & Units \\
                        & \multicolumn{1}{c}{{\sc spex}} & \multicolumn{1}{c}{{\sc xspec}} &  \\        
        \hline
        $kT_{1}$ & 0.49$_{-0.02}^{+0.02}$ & 0.53$_{-0.02}^{+0.02}$ & keV \\
        $kT_{2}$ & 1.02$_{-0.01}^{+0.02}$ & 1.06$_{-0.01}^{+0.01}$ & keV \\
        $kT_{3}$ & 3.15$_{-0.08}^{+0.08}$ & 3.18$_{-0.07}^{+0.07}$ & keV \\
        $N_{1}$ & 2.6$_{-0.2}^{+0.2}$ & 0.015$_{-0.001}^{+0.001}$ & 10$^{59}$~m$^{-3}$  \\
        $N_{2}$ & 6.6$_{-0.3}^{+0.3}$ & 0.038$_{-0.002}^{+0.002}$ & 10$^{59}$~m$^{-3}$  \\        
        $N_{3}$ & 15.54$_{-0.03}^{+0.03}$ & 0.085$_{-0.002}^{+0.002}$ & 10$^{59}$~m$^{-3}$  \\
        O & 0.74$_{-0.09}^{+0.09}$ & 0.43$_{-0.05}^{+0.06}$ & \\
        Ne & 1.11$_{-0.05}^{+0.05}$ & 1.23$_{-0.05}^{+0.05}$ & \\
        Na & 1.1$_{-0.2}^{+0.2}$ & 1.0$_{-0.2}^{+0.2}$ & \\
        Mg & 0.41$_{-0.02}^{+0.02}$ & 0.36$_{-0.02}^{+0.02}$ & \\
        Al & 0.36$_{-0.09}^{+0.09}$ & 0.34$_{-0.08}^{+0.09}$ & \\
        Si & 0.30$_{-0.01}^{+0.01}$ & 0.29$_{-0.01}^{+0.01}$ & \\
        S & 0.26$_{-0.04}^{+0.04}$ & 0.24$_{-0.03}^{+0.03}$ & \\
        Ar & 1.1$_{-0.1}^{+0.1}$ & 1.0$_{-0.1}^{+0.1}$ & \\        
        Ca & 0.8$_{-0.2}^{+0.2}$ & 0.8$_{-0.2}^{+0.2}$ & \\
        Fe & 0.28$_{-0.01}^{+0.01}$ & 0.180$_{-0.008}^{+0.008}$ & \\
        Ni & 0.20$_{-0.04}^{+0.04}$ & 0.15$_{-0.04}^{+0.04}$ & \\
        $z$ & 120.0$_{-3.8}^{+3.8}$ & 119.8$_{-3.6}^{+3.6}$ & km~s$^{-1}$ \\    
        C-statistics/d.o.f. & 1624.2/1350 & 1830.9/1362 & \\
        \hline
        \hline
    \end{tabular}\label{tab:sigmagem_spe}
\end{center}
\end{table}

\section{Discussion}
\label{sec:discussion}

In this paper, we extended our previous work on Capella (Paper\,I) to look for Doppler shifts in known stellar coronal sources that have been observed with the \chan\ ACIS-S operated in combination with the HETG. As summarized in Sect.~\ref{sec:intro}, stellar coronal sources are not only interesting for their relevance as Astrophysical laboratories of plasma physics and stellar evolution/interaction, but also because they are extensively used as calibration targets in the field of high energy resolution X-ray spectroscopy. This is mostly due to their emission features-rich spectra in the soft X-ray domain (typically $<$10~keV). The availability of multiple suitable calibration targets, beside Capella, is an advantage for present and future space-based observatories because calibrations can be tested over emission lines at different energies/intensities, and the availability of sources in different regions of the seasonally variable accessible sky can ease observational plans. 

Calibration observations targeting stellar coronal sources have been routinely performed for the gratings on-board \xmm\ and \chan\ in the past 24~years. The newer generation of X-ray instruments from space dedicated to high energy resolution observations, as the Resolve instrument on-board the \xrism\ mission launched in mid-2023 or the calorimeters on-board the future \athena\ and \lem\ missions later in the future, are aiming at calibrating via similar observations their energy resolution down to few eV \citep[at the moment of writing the expected launch for \athena\ is around 2037, while \lem\ is competing for a launch opportunity in 2032; see][and references therein]{xifu,lem}. For this reason, it is crucial to accurately measure if Doppler shifts exist in the binary systems leading to periodic changes in energy of the centroids of all emission lines. We showed in Paper\,I that Capella is a ``good'' calibration source because the Doppler shifts of the emission lines largely follow the orbit of the primary star and can thus be carefully corrected to improve the accuracy to be achieved for calibration purposes. Although the calibration plans for the X-IFU instrument and for the \lem\ mission are not yet fully detailed in publicly available documents, at least those spelled out for the Resolve instrument include additional stellar coronal sources beyond Capella \citep{xrismcal}. Our goal is to determine from the available HETG archive which of the systems in this class could be considered good calibration targets. We exploited our previously successful analysis techniques to a wider sample of stellar coronal sources, including eight objects with substantial orbital phase coverage from the HETG data and three with only limited coverage. 

In this context, our analysis show that, after Capella, HR\,1099 should probably be considered an equally good target for calibration observations, as the HETG data demonstrated that the velocity shifts in this source closely follow the orbit of the primary star. Most of the data points used for this analysis are coming from a single HETG observation (ObsID~62538) but the fact that also the data point obtained from the ObsID~1252, carried out 3 days after the first observation, is lying accurately along the same line adds confidence that radial velocity shifts in this system are largely predictable. Additional HETG data covering at least the portions of the orbit left uncovered before (0.0-0.1 and 0.6-1.0) and carried out at present times could be used to strengthen this conclusion and confirm HR\,1099 is a reliable calibration target. 

IM\,Peg also proved to be a promising target. Compared to the case of HR\,1099, the HETG data covered more uniformly the orbital phases and  yet the {\sc spex} and {\sc xspec} measurements aligned well along the line of the radial velocities expected for the primary star. The data used for the analysis spanned a total of $\sim$45~days, thus covering about two orbital periods of the source. The fact that the measurements from two different orbits align well along the same prediction adds confidence that the velocity shifts in IM\,Peg are accurately predictable. 

Potentially interesting targets requiring further investigations and definitively additional data are UX\,Ari, $\sigma$\,Gem, and HR\,5110. In the case of UX\,Ari, the HETG results showed that large velocity shifts are measured close to the values expected along the orbit of the primary star. The difference of about 20\% between the measured HETG Doppler shifts and those predicted along the orbit of the primary star led us to suggest that some activity from the secondary stellar component could affect the results and lead to lower velocity shifts. Our simulations showed that a contamination at the level of 10\% from the secondary star at this orbital phase would be already capable of producing the measured reduction in the Doppler shifts. On one hand, UX\,Ari would thus be a desirable target because of the large velocity shifts in the system and the possibility that HETG measurements closely follow the orbit of the primary star. On the other hand, the data we have available today suggest that a contamination from the secondary star cannot be excluded and this  would make the variations of the Doppler shift along the orbit less accurately predictable compared to more favorable targets. More data are definitively needed for this object, as so far only $\sim$0.1 in orbit has been covered by the HETG data. 

In the case of $\sigma$\,Gem, the HETG data covers as of today only a limited portion of the orbit ($\lesssim$0.1) but the obtained measurements from the {\sc spex} and {\sc xspec} fits strongly resemble the findings discussed for HR\,1099 and IM\,Peg, with the data points closely matching the line of the radial velocities expected along the orbit of the primary star. Velocities are also among the highest in the considered stellar coronal sources and thus we suggest that additional HETG data are strongly needed in order to confirm the apparent suitability of $\sigma$\,Gem as a calibration target. 

In the case of HR\,5110, we also found indications that the HETG measurements could align along the orbit of the primary, but the uncertainties on the current measurements are far too large to draw a firm conclusion. Additional HETG observations could help clarifying our findings but observations should have a significantly longer exposure (at least a factor of 2) in order to reduce the error bars. Although this would sum up to a substantial observational time, the relatively short orbital period of the system (2.61~d) would ease the achievement of monitoring (at least) one complete orbital revolution. 

We classify as less promising sources for calibration purposes the remaining three targets, i.e. AR\,Lac, V824\,Ara, and TZ\,CrB. In all these systems, our analysis could not reveal a clear trend of the Doppler shift measurements along either the orbit of the primary or secondary star and fluctuations are clearly less predictable (and thus more difficult to be corrected for) than in the other sources discussed above (despite being sometimes as large as $\simeq$50~km~s$^{-1}$). Among these three objects, TZ\,CrB shows the least fluctuating Doppler velocities, with the average of the Doppler shifts being reasonably close to zero. Although one could argue that this situation is rather favorable for calibration activities as practically no correction is needed, we remark that the HETG observation available so far is covering a single orbit and thus we cannot exclude that fluctuations will increase when more orbits are taken into account. A similar conclusion applies to the case of V824\,Ara, while fluctuations of the Doppler velocities in an unpredictable manner seem definitively larger in the case of AR\,Lac. It is interesting to note that our results for these three systems do not come completely unexpected. Literature papers have discussed how all these three systems are characterized by the presence of both a primary and secondary stellar component which are chromospherically active, thus leading to contributions to the Doppler shifts which can interfere. It has to be remarked that both V824\,Ara and TZ\,CrB present further complications for calibrations of future instruments, as these are known to be part of multiple hierarchical systems and the additional components beside the main inner binary are also relatively bright X-ray sources. In the case of V824\,Ara, we were able to disentangle the contribution of the inner chromospherically active binary from the contribution of the M3 star, but in the case of TZ\,CrB not even the excellent spatial resolution of \chan\ could be used to avoid the contamination of the emission from $\sigma$1\,CrB into the dispersed spectra of the source of interest. Avoiding contaminations in these sources could thus be even more complicated for future instruments which are not endowed with the same (or better) spatial resolution than \chan.\  

For three additional targets, i.e. $\lambda$\,And, II\,Peg, and Ty\,Pyx, the results of our analysis are only provided in the Appendix. For all these systems, the available HETG data do not allow us to reach firm conclusions about the measured versus predicted Doppler shifts. In both $\lambda$\,And and II\,Peg, only the radial velocities of the primary stars are known, and the measurements obtained from the HETG data do not seem to align too well with the expected values (although the mismatch between observations and expected values is much less for II\,Peg; see Fig.~\ref{fig:lambdaand_orbit} and \ref{fig:iipeg_orbit}). In the case of Ty\,Pyx, the situation is different because the sole HETG observation carried out as of today was centered at conjunction, which, by definition, is not the optimal orbital phase to disentangle between the orbit of the primary and secondary star. However, the quantitatively good match between the Doppler shift in the HETG data and the value expected according to the radial velocities published by\citet{andersen75} suggest that this source should be better investigated in the future via additional HETG observational campaigns to revise its possible role as good calibration target.

\section{Conclusion}
\label{sec:conclusion}

As discussed in our Paper\,I, Capella is so far the best stellar coronal source to carry out calibration observations of X-ray instruments dedicated to high resolution spectroscopy. This is mainly because the Doppler velocity shifts in this source accurately follow the radial velocities expected along the orbit of the primary star, and also the spread of the shifts around the predicted radial velocity curve could be measured over a time frame of over 22~years thus solidly proving its repeatability and stability over time. None of the other stellar coronal sources observed by the HETG on-board \chan\ have a similarly long baseline of observations. 

Thanks to our new analyses, we identified in this paper additional promising candidates to be used as calibration targets, mainly based on the evidence that the Doppler shifts  measured by the HETG on-board \chan\ closely follow the radial velocities expected along the orbit of one of the two stars hosted inside these systems. HR\,1099 and IM\,Peg have been classified as the most probable usable targets thanks to the close match of the HETG measured Doppler shifts with the radial velocities expected along the orbit of their primary star component. Compared to Capella, we are missing a test of these matches on the long term, as the available observations cover at best a few orbital periods. If it is assumed that the Doppler shifts are closely repeatable across different orbits, systems as Capella, HR\,1099, and IM\,Peg, allow us to reach confidence about the energy centroid of the emission lines down to an accuracy of typically $<$1~km~s$^{-1}$. This is the usual uncertainty of measurements obtained via observations in the optical, which constitute as of today our most accurate measurement of the radial velocities in coronal stellar sources (see also the discussion in Paper\,I). All other analyzed targets showed evidences of large scatters in the HETG measurements and the lack of a convincing alignment between the HETG measured Doppler shifts and the radial velocities predicted along the orbits of either the primary or the secondary stellar component. In most cases, this can be likely ascribed to the fact that, contrary to Capella, HR\,1099, and IM\,Peg, in these other systems both stellar components are active in the X-ray domain. Their contributions to the total X-ray emission leads to variations of the Doppler shifts by up to $\sim$50~km~s$^{-1}$ that can hardly be predicted and accounted for. Given the fact that in the X-ray domain the contributions of the primary and secondary stars to the total X-ray emission cannot be disentangled, we suggested that these systems cannot be considered ``good'' calibration targets. In most sources, the lack of a long-term observational campaign also hampers any attempt to verify if the Doppler shifts are repeatable or not across different orbital revolutions. We conclude that for all these systems, it is unlikely that we can control intrinsic systematics to a better accuracy than $\sim$50~km~s$^{-1}$. Furthermore, we probed with our analyses that results for the different sources in terms of Doppler velocity shifts obtained within the {\sc spex} and {\sc xspec} spectral fit environments are broadly compatible to within the associated 1~$\sigma$ c.l. uncertainties, albeit the two environments exploit independent atomic databases.

\section*{Data availability}
All data exploited in this paper are publicly available from the \chan\ archive and processed with publicly available software.

\section*{Acknowledgements}
We thank the anonymous referee for useful comments. EB thanks SRON for the kind hospitality during part of this work. This paper is supported by European Union's 2020 research and innovation programme under grant agreement No.871158, project AHEAD2020. Support for DPH was provided by NASA through the Smithsonian Astrophysical Observatory (SAO) contract SV3-73016 to MIT for Support of the Chandra X-Ray Center (CXC) and Science Instruments. CXC is operated by SAO for and on behalf of NASA under contract NAS8-03060.

\bibliography{bib}{}
\bibliographystyle{mnras}

\appendix

\section{Source with limited orbital phase coverage} 

We report here the results that we obtained for three of the analyzed sources that had only a limited orbital phase coverage 
through all the HETG observations available so far. For these sources, the same analysis techniques used for the sources in the main paper are exploited, but we could not achieve a firm conclusion about the presence of Doppler shifts within the binaries following the orbit of one of the two stellar components (or none of them). Results here are thus presented mainly for completeness. 

\subsection{$\lambda$\,And} 
\label{sec:lambdaand}

$\lambda$\,And is a single-lined spectroscopic binary system hosting a G8III-IV star and it belongs to the 
class of RS CVn variables \citep[see, e.g.,][and references therein]{martinez21}. The estimated  
distance to the source is 26.41~pc \citep{gaia3}. \citet{walker44} reported the complete ephemerides 
of the system, measuring an orbital period of 
20.5212~days and a relatively modest eccentricity of 0.040-0.084 (see their Table~IV).  
Although updated ephemerides were reported in two more recent papers \citep{eaton07,massarotti08}, there were no improvements in the accuracy of the determined orbital period compared to \citet{walker44}. 
 \begin{figure}
  \centering
  \includegraphics[width=9.0cm]{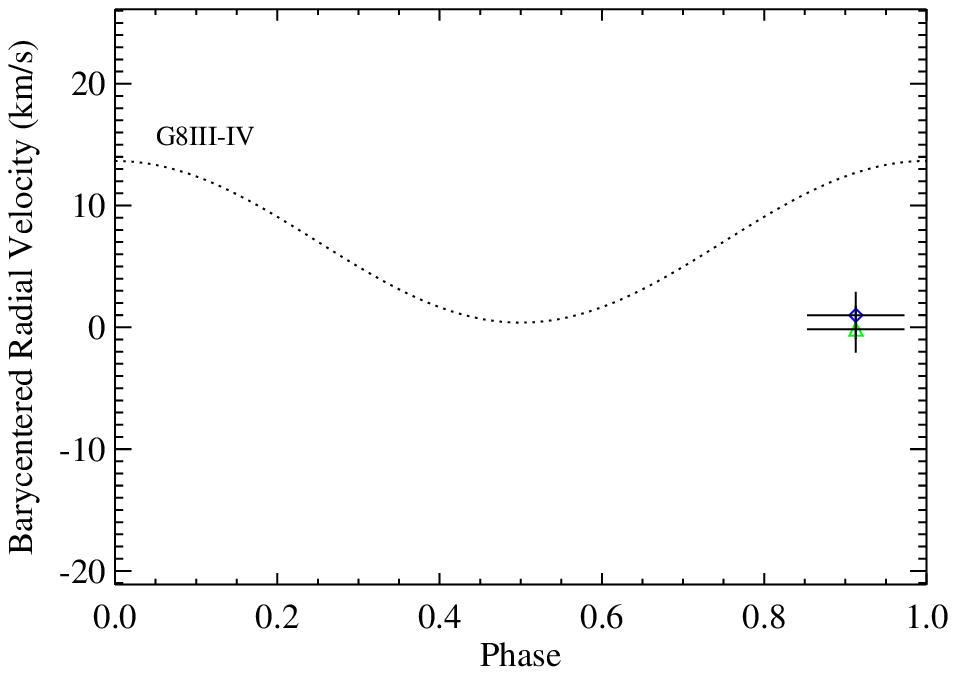}
  \caption{\label{fig:lambdaand_orbit} Same as Fig.~\ref{fig:hr1099_orbit} but for the case of $\lambda$\,And. The dotted line represents the expected barycentered radial velocities of the G8\,III-IV star calculated according to the ephemerides published by \citet{walker44}. The green point corresponds to value obtained from the spectral fits in {\sc spex}, while the blue point  was obtained from the fits in {\sc xspec}. The error bars for the radial velocity measurement are given at 1~$\sigma$~c.l. and are only statistical. The uncertainty on the phase includes the duration of the observation as well as the uncertainties on the ephemerides indicated by \citet{walker44}.}  
\end{figure}

High resolution spectroscopy of $\lambda$\,And was presented by \citet{audard03} using \xmm/RGS data. These authors found that the spectrum of the source could be well described by using a model comprising 4 CIE components, out of which one at particularly high energy ($\gtrsim$5~keV) compared to other similar systems (see their Table~4). The source was also observed once (ObsID~609) by \chan\ using the combination of the ACIS-S with the HETG for a total exposure time of about 82~ks (see Table~\ref{tab:lambdaand_obs}). The sourced displayed a fairly constant X-ray luminosity across the observation \citep{forcada04}. These data were exploited in a number of previous publications to analyze specific emission lines and investigate both the density and the X-ray optical depth of coronal plasma \citep[see][]{testa04,testa04b,ness04,testa07}. 

As done for all other sources in this paper, we first extracted the MEG +1 spectrum using the entire exposure time available within the ObsID~609. Following \citet{audard03}, we attempted a fit with 4 CIE components in {\sc spex} but found that the fit was insensitive to the highest energy component. Therefore, we used for our best fit results a model comprising only 3 CIE components (see Fig.~\ref{fig:lambdaand_plot_spex}). We included in the fit a REDS component to take into account the velocity shifts of the lines within the binary and verified that the addition of an HOT component to mimic the effect of the Galactic absorption did not significantly affect the fit results \citep[given the low value of the absoprtion column density, see][]{ortolani97,pandey12}. 
The best fit parameters obtained from the MEG +1 spectrum are reported in Table~\ref{tab:lambdaand_spe}. Given the relatively long orbital period of the source and the know low expected radial velocities in the system, we did not split the HETG observation in sub-segments (the data provided a limited coverage of $\pm$0.023 in orbital phase). We measured an overall velocity shift by fitting the remaining HETG spectra (MEG -1, HEG $\pm$1) with the same spectral model described above (fixing the element abundances and the temperatures of the CIE components) and computing a weighted average of the four measured velocity shifts. The results are reported in Table~\ref{tab:lambdaand_obs} and displayed in Fig.~\ref{fig:lambdaand_orbit}. 
In all these tables and figures, we also reported as for the previous sources the results obtained from the fits of the HETG spectra within {\sc xspec} using the same spectral model described above. Note that the large error bar associated to the phase of the ObsID~609 in Fig.~\ref{fig:lambdaand_orbit} is not related to the coverage in phase of the data but mostly to the uncertainty reported on the reference time in \citet{walker44}.
\begin{table*}
\caption{Same as Table~\ref{tab:hr1099_obs} but for the \chan/HETG observation of $\lambda$\,And. The orbital phase is calculated from the ephemerides published by \citet[we used the D.A.O. solution, see their Table~IV;][]{walker44}. All indicated uncertainties are at 1$\sigma$~c.l. The uncertainty on the phase includes the known uncertainties on the published source ephemerides.} 
\label{tab:lambdaand_obs}
\begin{tabular}{ccccccc}
\hline
 & Middle & Effective &   & Corrected velocity & Corrected velocity & Barycenter correction \\
 & Observational time & exposure & Phase & V$_{{\rm corr}_{\rm spex}}$ & V$_{{\rm corr}_{\rm xspec}}$ & V$_{\rm bary}$ \\
ObsID & (HJD)  & (ksec)   & $\phi$ & (km/s) & (km/s) & (km/s) \\
\hline
609  & 2451530.1687 &  16.2 & 0.91$_{-0.04}^{+0.04}$ & -0.2$_{-1.9}^{+1.9}$ & 1.0$_{-1.9}^{+1.9}$ &  20.2 \\ 
\hline
\end{tabular}
\end{table*}
\begin{table}
    \begin{center}
    \caption{\label{tab:lambdaand_spe} Same as Table~\ref{tab:hr1099_spe}, but for the case of the +1 order MEG spectrum of $\lambda$\,And obtained from the ObsID~609. The best fit was obtained by using three plasma components (CIE in {\sc spex} and BVVAPEC in {\sc xspec}). The fit in {\sc spex} also includes a REDS component.}
    \begin{tabular}{llll}
        \hline
        \hline
        Model parameter & \multicolumn{1}{c}{Best fit values} & \multicolumn{1}{c}{Best fit values} & Units \\
                        & \multicolumn{1}{c}{{\sc spex}} & \multicolumn{1}{c}{{\sc xspec}} &  \\        
        \hline
        $kT_{1}$ & 0.44$_{-0.02}^{+0.02}$ & 0.51$_{-0.02}^{+0.02}$ & keV \\
        $kT_{2}$ & 0.84$_{-0.02}^{+0.02}$ & 0.90$_{-0.02}^{+0.02}$ & keV \\
        $kT_{3}$ & 1.46$_{-0.05}^{+0.05}$ & 1.51$_{-0.06}^{+0.06}$ & keV \\
        $N_{1}$ & 0.73$_{-0.08}^{+0.08}$ & 0.010$_{-0.001}^{+0.001}$ & 10$^{59}$~m$^{-3}$  \\
        $N_{2}$ & 1.83$_{-0.09}^{+0.09}$ & 0.021$_{-0.001}^{+0.001}$ & 10$^{59}$~m$^{-3}$  \\        
        $N_{3}$ & 1.12$_{-0.08}^{+0.0.09}$ & 0.012$_{-0.001}^{+0.001}$ & 10$^{59}$~m$^{-3}$  \\
        O & 0.56$_{-0.07}^{+0.07}$ & 0.35$_{-0.04}^{+0.05}$ & \\
        Ne & 0.68$_{-0.03}^{+0.03}$ & 0.73$_{-0.04}^{+0.04}$ & \\
        Na & 0.4$_{-0.1}^{+0.1}$ & 0.4$_{-0.1}^{+0.1}$ & \\
        Mg & 0.50$_{-0.02}^{+0.02}$ & 0.47$_{-0.02}^{+0.02}$ & \\
        Al & 0.61$_{-0.07}^{+0.08}$ & 0.58$_{-0.07}^{+0.07}$ & \\
        Si & 0.23$_{-0.01}^{+0.01}$ & 0.23$_{-0.03}^{+0.03}$ & \\
        S & 0.15$_{-0.03}^{+0.03}$ & 0.14$_{-0.03}^{+0.03}$ & \\
        Ca & 0.9$_{-0.3}^{+0.4}$ & 0.6$_{-0.3}^{+0.3}$ & \\
        Fe & 0.167$_{-0.008}^{+0.009}$ & 0.102$_{-0.005}^{+0.005}$ & \\
        Ni & 0.13$_{-0.02}^{+0.03}$ & 0.14$_{-0.03}^{+0.03}$ & \\
        $z$ & 47.4$_{-3.6}^{+3.6}$ & 48.3$_{-3.6}^{+3.6}$ & km~s$^{-1}$ \\    
        C-statistics/d.o.f. & 1614.5/1237 & 1799.2/1217 & \\
        \hline
        \hline
    \end{tabular}
\end{center}
\end{table}

From this figure, we note that the obtained velocity shift values from our X-ray analysis are somewhat lower than those predicted along the orbit of the G8\,III-IV star. This does not come unexpected as the overall radial velocities in the system are particularly low compared to other sources. We thus argue that radial velocity measurements in $\lambda$\,And are particularly challenging compared to other sources of the same class and at the limit of the capabilities of high resolution X-ray instruments on-board currently operating missions. Furthermore, the companion of the G8\,III-IV star and its orbits are not known, so it is currently not possible to evaluate the effect of this star component on the measured radial velocities.

\subsection{II\,Peg} 
\label{sec:iipeg}

II\,Peg is a single-lined spectroscopic binary, and the sole star known in the system (the primary) is classified as a K2\,IV subgiant  at 39.36~pc \citep[see, e.g.,][and references therein]{xiang14,gaia2}. The most recent ephemerides of the system were published by \citet{rosen15}, reporting also the updated curve of the measured radial velocities along the orbit of the primary. 

II\,Peg is known to be a particularly active binary and it has been observed several times in X-rays. At low resolution, \citet{covino00} studied the source broad-band spectrum as observed with the instruments on-board \beppo\ and found that the emission could be described by a model comprising two thermal plasma components and an absorption component with a measured column density of 7.8$\times$10$^{22}$~cm$^{-2}$. At high resolution, II\,Peg was observed only once with the ACIS-S on-board \chan\ in combination with the HETG (ObsID~1451). These data have been studied in details using a line-based analysis by \citet{dph01} and subsequently exploited also to investigate both the density and the X-ray optical depth of coronal plasmas \citep{testa04,testa04b,ness04,testa07}. During the ObsID.~1451, the sourced displayed an X-ray flare which onset was recorded about 25~ks after the beginning of the exposure and covered the remaining part of the observation \citep{dph01}.
\begin{table*}
\caption{Same as Table~\ref{tab:hr1099_obs} but for the \chan/HETG observation of II\,Peg. The orbital phase is calculated from the ephemerides published by \citet{rosen15}. All indicated uncertainties are at 1$\sigma$~c.l. The uncertainty on the phase includes the known uncertainties on the published source ephemerides, which are largely dominating the resulting size of the error bars.} 
\label{tab:iipeg_obs}
\begin{tabular}{ccccccc}
\hline
 & Middle & Effective &   & Corrected velocity & Corrected velocity & Barycenter correction \\
 & Observational time & exposure & Phase & V$_{{\rm corr}_{\rm spex}}$ & V$_{{\rm corr}_{\rm xspec}}$ & V$_{\rm bary}$ \\
ObsID & (HJD)  & (ksec)   & $\phi$ & (km/s) & (km/s) & (km/s) \\
\hline
1415  & 2454562.9719 &  43.3 & 0.87$_{-0.19}^{+0.19}$ & -10.3$_{-2.6}^{+2.6}$ & -3.4$_{-2.6}^{+2.6}$ &  6.6 \\ 
\hline
\end{tabular}
\end{table*}
\begin{table}
    \begin{center}
\label{tab:iipeg_spe}
    \caption{Same as Table~\ref{tab:hr1099_spe}, but for the case of the +1 order MEG spectrum of II\,Peg obtained from the ObsID~1415. The best fit was obtained by using three plasma components (CIE in {\sc spex} and BVVAPEC in {\sc xspec}). The fit in {\sc spex} also includes a REDS component.}
    \begin{tabular}{llll}
        \hline
        \hline
        Model parameter & \multicolumn{1}{c}{Best fit values} & \multicolumn{1}{c}{Best fit values} & Units \\
                        & \multicolumn{1}{c}{{\sc spex}} & \multicolumn{1}{c}{{\sc xspec}} &  \\        
        \hline
        $kT_{1}$ & 0.40$_{-0.02}^{+0.02}$ & 0.44$_{-0.02}^{+0.02}$ & keV \\
        $kT_{2}$ & 1.01$_{-0.03}^{+0.03}$ & 1.04$_{-0.04}^{+0.03}$ & keV \\
        $kT_{3}$ & 3.0$_{-0.1}^{+0.1}$ & 3.0$_{-0.1}^{+0.1}$ & keV \\
        $N_{1}$ & 1.3$_{-0.2}^{+0.2}$ & 0.007$_{-0.001}^{+0.001}$ & 10$^{59}$~m$^{-3}$  \\
        $N_{2}$ & 3.3$_{-0.3}^{+0.4}$ & 0.017$_{-0.002}^{+0.002}$ & 10$^{59}$~m$^{-3}$  \\        
        $N_{3}$ & 11.23$_{-0.03}^{+0.03}$ & 0.060$_{-0.002}^{+0.002}$ & 10$^{59}$~m$^{-3}$  \\
        O & 1.5$_{-0.1}^{+0.2}$ & 0.9$_{-0.1}^{+0.1}$ & \\
        Ne & 1.6$_{-0.1}^{+0.1}$ & 1.7$_{-0.1}^{+0.1}$ & \\
        Mg & 0.36$_{-0.03}^{+0.03}$ & 0.34$_{-0.03}^{+0.03}$ & \\
        Al & 0.6$_{-0.2}^{+0.2}$ & 0.7$_{-0.2}^{+0.2}$ & \\
        Si & 0.27$_{-0.02}^{+0.02}$ & 0.27$_{-0.02}^{+0.02}$ & \\
        S & 0.32$_{-0.05}^{+0.06}$ & 0.29$_{-0.05}^{+0.05}$ & \\
        Ca & 0.5$_{-0.2}^{+0.3}$ & 0.5$_{-0.2}^{+0.2}$ & \\
        Fe & 0.17$_{-0.01}^{+0.01}$ & 0.111$_{-0.008}^{+0.009}$ & \\
        Ni & 0.12$_{-0.06}^{+0.07}$ & 0.13$_{-0.06}^{+0.07}$ & \\
        $z$ & 14.1$_{-5.1}^{+5.1}$ & 19.8$_{-4.8}^{+4.8}$ & km~s$^{-1}$ \\    
        C-statistics/d.o.f. & 1435.5/1320 & 1438.7/1313 & \\
        \hline
        \hline
    \end{tabular}
\end{center}
\end{table}

\begin{figure}
  \centering
  \includegraphics[width=9.0cm]{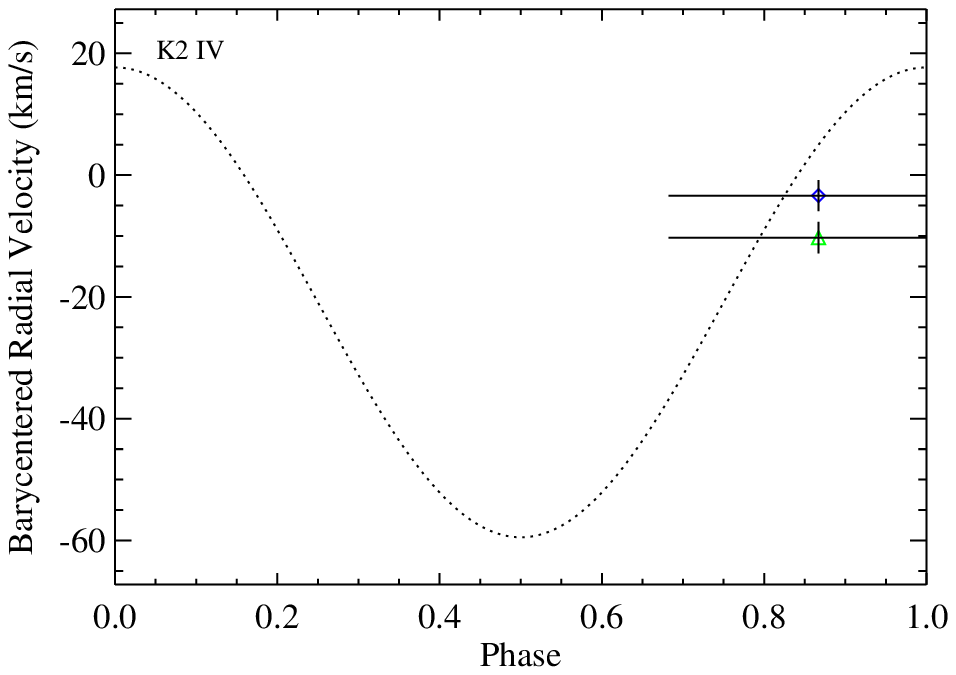}
  \caption{\label{fig:iipeg_orbit} Same as Fig.~\ref{fig:hr1099_orbit} but for the case of II\,Peg. The dotted line represents the expected barycentered radial velocities of the primary star calculated according to the ephemerides published by \citet{rosen15}. The green point corresponds to value obtained from the spectral fits in {\sc spex}, while the blue point  was obtained from the fits in {\sc xspec}. The error bars for the radial velocity measurement are given at 1~$\sigma$~c.l. and are only statistical. The uncertainty on the phase includes the duration of the observation as well as the uncertainties on the ephemerides.}  
\end{figure}

Here, we re-analyzed the ACIS-S HETG observation of II\,Peg using the same techniques we already illustrated for all other sources. We first extracted the MEG +1 spectrum and attempted a fit within {\sc spex} using a model comprising 3 CIE components. We included in the fit a HOT component to mimic the effect of the Galactic absorption with a value fixed at 7.8$\times$10$^{22}$~cm$^{-2}$ and a REDS component to evaluate velocity shifts within the binary. The model gave an acceptable fit to the spectrum of the source, leaving no lines unaccounted for in the residuals (see Fig.~\ref{fig:iipeg_plot_spex}). We also refit the same spectrum in {\sc xspec} using an equivalent model (i.e. adopting 3 BVVAPEC components plus the Galactic absorption) to compare the measured velocity shifts in the two environments. The results of this analysis are reported in Table~\ref{tab:iipeg_spe}. We then applied the same spectral model described above to fit the remaining MEG -1 and HEG $\pm$1 spectra extracted from the ObsID~1451. The velocity shift within the binary corresponding to this observation was obtained by using a weighted average of all results from the four spectra separately within {\sc spex} and {\sc xspec}. We summarize our findings in Fig.~\ref{fig:iipeg_orbit} and in Table~\ref{tab:iipeg_obs}. This table also reports the computed orbital phase of the ObsID~1415 when the ephemerides from \citet{rosen15} are considered. Note that the large error bar associated to the phase of the ObsID~1415 in Fig.~\ref{fig:iipeg_orbit} is not related to the coverage in phase of the data but mostly to the uncertainty reported on the reference time in \citet{rosen15}. As the orbital period of the system is about 6.4~days and the ObsID~1415 is only 43.3~ks long, the data span a limited orbital phase coverage ($\lesssim$0.08). For this reason, we did not attempt to split the observation in more segments as done for other sources. 

From the results in Fig.~\ref{fig:iipeg_orbit} we suggest that the single velocity shift measurement available from the \chan\ HETG data could be in agreement with those expected along the orbit of the primary, although there is definitively the need of additional observations to draw a firm conclusion.

\subsection{Ty\,Pyx} 
\label{sec:typyx}

Ty\,Pyx is a chromospherically active binary hosting two star components of spectral type G5\,IV and showing relatively sharp eclipses \citep[see, e.g.,][and references therein]{strassmeier09,wilson10}. The system orbital period is estimated at 3.2~days and the measured distance is 56~pc \citep{leeuwen07}. At the best of our knowledge, the most recent publication in the literature reporting the ephemerides of the system as well as the measured radial velocities of the two stellar component is by \citet{andersen75}. These authors provided a detailed estimate of all relevant geometrical system parameters and, although the uncertainties on their reported values of the reference time and orbital period were not explicitly indicated, they commented that their analysis could achieve an accuracy on the orbital phase determination of about 1\%. 

In the X-ray domain, Ty\,Pyx was observed at both high and low spectral resolution. By using \beppo\ data, \citet{franciosini03} provided a fairly accurate description of the source broad-band X-ray spectrum, including an estimate of the absorption column density in the direction of the source at 1.2$\times$10$^{19}$~cm$^{-2}$ \citep[see also][]{culhane90,huenemoerder98}. At high spectral resolution, the source was observed with both the \chan\ and \xmm\ grating instruments. At the best of our knowledge, the \xmm\ data remains to date yet to be published, while the \chan\ data have been only preliminary reported by \citet[showing only the unfitted HETG spectrum;][]{huenemoerder98} and subsequently used for the study of a few specific lines with the aim of investigating the size/density of coronal plasmas and the coronae X-ray optical depth in a large sample of chromospherically active binaries \citep{ness03,ness04,testa04,testa07}.  

\begin{figure}
  \centering
  \includegraphics[width=9.0cm]{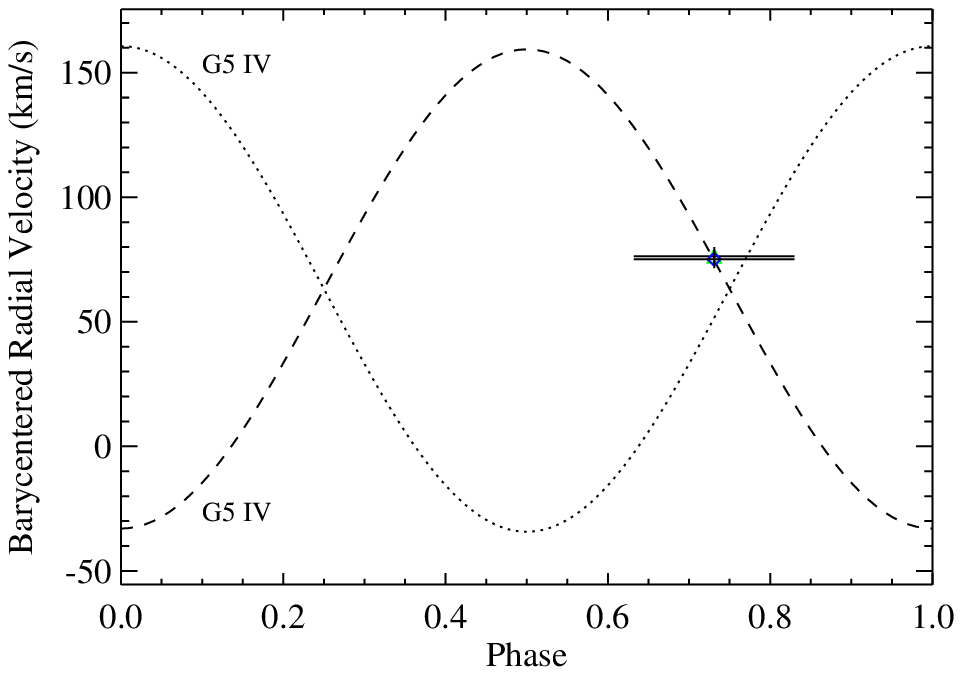}
  \caption{\label{fig:typyx_orbit} Same as Fig.~\ref{fig:hr1099_orbit} but for the case of Ty\,Pyx. The dotted line represents the expected barycentered radial velocities of the primary star calculated according to the ephemerides published by \citet{andersen75}. The green point corresponds to value obtained from the spectral fits in {\sc spex}, while the blue point  was obtained from the fits in {\sc xspec}. The error bars for the radial velocity measurement are given at 1~$\sigma$~c.l. and are only statistical. The uncertainty on the phase includes the duration of the observation as well as the uncertainties on the ephemerides.}  
\end{figure}

Here, we re-analyzed the sole available \chan\ ACIS-S observation of Ty\,Pyx carried out in combination with the HETG (ObsID~601). During this observation, the source displayed a relatively moderate flare occurring roughly around the middle of the exposure \citep{huenemoerder02}. By following the same analysis steps as all other sources, we first fit the MEG +1 spectrum of the source with a model comprising three plasma components in both the {\sc spex} and {\sc xspec} environments. As done before, we included in the {\sc spex} fit a REDS component to take into account velocity shifts in the binary and in both environments a component to mimic the effect of the Galactic absorption (specifically, a HOT component in {\sc spex} and a TBABS component in {\sc xspec}). The absorption column density in the direction of the source was kept fixed in all fits to the value of 1.2$\times$10$^{19}$~cm$^{-2}$ reported previously in the literature (see above). We obtained both for the {\sc spex} and {\sc xspec} cases satisfactorily fits, with no evident structured residuals around any of the detected emission lines. The best fit parameters are summarized in Table~\ref{tab:typyx_spe} (see also Fig.~\ref{fig:typyx_plot_spex}). The same model was able to satisfactorily describe also the remaining HETG spectra from this observation, i.e. the MEG -1 and the HEG $\pm$1 spectra. We obtained an estimate of the velocity shift in the binary in both the {\sc spex} and {\sc xspec} environments by weighted-averaging the values obtained from the individual fits to the four available spectra. We report the results of this analysis in Fig.~\ref{fig:typyx_orbit} and in Table~\ref{tab:typyx_obs}.Values obtained from the {\sc spex} and {\sc xspec} environments turned out to be well compatible to within the estimated 1~$\sigma$ uncertainties. Table~\ref{tab:typyx_obs} also reports the estimated orbital phase and orbital coverage corresponding to the ObsID~601 calculated using the ephemerides from \citet{andersen75}. We note from Fig.~\ref{fig:typyx_orbit} that the measured velocity shifts are well compatible with the value expected at the inferior conjunction. By definition, this orbital phase is not suited to discriminate if the velocity shifts are more associated to the X-ray activity from the primary or the secondary star, and thus lacking further observations it is difficult to achieve any conclusion. Given the fact that the data span a non-negligible part of the binary orbit (about 0.18), we also attempted to split the ObsID~601 in two separated segments. However, the measured orbital shifts differed by only $\sim$3~km~s$^{-1}$. This is well within the uncertainties affecting these measurements ($\pm$4~km~s$^{-1}$) and in any case in line with what expected at this orbital phase. 
\begin{table*}
\caption{Same as Table~\ref{tab:hr1099_obs} but for the \chan/HETG observation of Ty\,Pyx. The orbital phase is calculated from the ephemerides published by \citet{andersen75}. All indicated uncertainties are at 1$\sigma$~c.l. The uncertainty on the phase includes the known uncertainties on the published source ephemerides.} 
\label{tab:typyx_obs}
\begin{tabular}{ccccccc}
\hline
 & Middle & Effective &   & Corrected velocity & Corrected velocity & Barycenter correction \\
 & Observational time & exposure & Phase & V$_{{\rm corr}_{\rm spex}}$ & V$_{{\rm corr}_{\rm xspec}}$ & V$_{\rm bary}$ \\
ObsID & (HJD)  & (ksec)   & $\phi$ & (km/s) & (km/s) & (km/s) \\
\hline
601  & 2451912.9049 &  49.1 & 0.48$_{-0.09}^{+0.09}$ & 76.4$_{-3.6}^{+3.6}$ & 75.1$_{-3.6}^{+3.6}$ &  -15.9 \\ 
\hline
\end{tabular}
\end{table*}
\begin{table}
    \begin{center}
\label{tab:typyx_spe}
    \caption{Same as Table~\ref{tab:hr1099_spe}, but for the case of the +1 order MEG spectrum of Ty\,Pyx obtained from the ObsID~601. The best fit was obtained by using three plasma components (CIE in {\sc spex} and BVVAPEC in {\sc xspec}). The fit in {\sc spex} also includes a REDS component.}
    \begin{tabular}{llll}
        \hline
        \hline
        Model parameter & \multicolumn{1}{c}{Best fit values} & \multicolumn{1}{c}{Best fit values} & Units \\
                        & \multicolumn{1}{c}{{\sc spex}} & \multicolumn{1}{c}{{\sc xspec}} &  \\        
        \hline
        $kT_{1}$ & 0.41$_{-0.02}^{+0.02}$ & 0.44$_{-0.02}^{+0.03}$  & keV \\
        $kT_{2}$ & 0.90$_{-0.03}^{+0.03}$ & 0.90$_{-0.03}^{+0.04}$  & keV \\
        $kT_{3}$ & 1.8$_{-0.1}^{+0.1}$ & 1.9$_{-0.1}^{+0.1}$  & keV \\
        $N_{1}$ & 1.0$_{-0.1}^{+0.1}$ & 0.0027$_{-0.0004}^{+0.0004}$ & 10$^{59}$~m$^{-3}$  \\
        $N_{2}$ & 1.5$_{-0.1}^{+0.1}$ & 0.0042$_{-0.0004}^{+0.0004}$ & 10$^{59}$~m$^{-3}$  \\        
        $N_{3}$ & 2.8$_{-0.2}^{+0.2}$ & 0.0076$_{-0.0005}^{+0.0005}$ & 10$^{59}$~m$^{-3}$  \\
        O & 0.9$_{-0.2}^{+0.3}$ & 0.5$_{-0.1}^{+0.1}$ & \\
        Ne & 1.2$_{-0.1}^{+0.1}$ & 1.3$_{-0.2}^{+0.2}$ & \\
        Na & 2.8$_{-0.7}^{+0.8}$ & 2.4$_{-0.6}^{+0.7}$ & \\
        Mg & 0.74$_{-0.08}^{+0.10}$ & 0.65$_{-0.08}^{+0.08}$ & \\
        Al & 1.1$_{-0.3}^{+0.3}$ & 1.1$_{-0.3}^{+0.3}$ & \\
        Si & 0.51$_{-0.05}^{+0.06}$ & 0.49$_{-0.05}^{+0.05}$ & \\
        S & 0.5$_{-0.1}^{+0.1}$ & 0.4$_{-0.1}^{+0.1}$ & \\
        Ar & 0.2$_{-0.2}^{+0.3}$ & 0.2$_{-0.2}^{+0.3}$ & \\        
        Ca & 1.6$_{-0.7}^{+0.8}$ & 1.5$_{-0.6}^{+0.7}$ & \\
        Fe & 0.58$_{-0.05}^{+0.06}$ & 0.36$_{-0.04}^{+0.04}$ & \\
        Ni & 0.5$_{-0.1}^{+0.1}$ & 0.5$_{-0.1}^{+0.1}$ & \\
        $z$ & 75.8$_{-7.9}^{+7.9}$ & 75.4$_{-7.7}^{+7.7}$ & km~s$^{-1}$ \\    
        C-statistics/d.o.f. & 1224.6/1034 & 1129.4/938 & \\
        \hline
        \hline
    \end{tabular}
\end{center}
\end{table}

\clearpage

\section{Spectral fit plots} 
\label{sec:plots}

We report for completeness in this section the figures corresponding to the fits of the average source spectra within the {\sc spex} and {\sc xspec} environments. The plots below correspond in order of appearance to spectral fit results displayed in Table~\ref{tab:hr1099_spe}, \ref{tab:impeg_spe}, \ref{tab:arlac_spe}, \ref{tab:uxari_spe}, \ref{tab:v824ara_spe}, \ref{tab:tzcrb_spe}, \ref{tab:hr5110_spe}, \ref{tab:sigmagem_spe}, \ref{tab:lambdaand_spe}, \ref{tab:iipeg_spe}, and \ref{tab:typyx_spe}.   
 \begin{figure*}
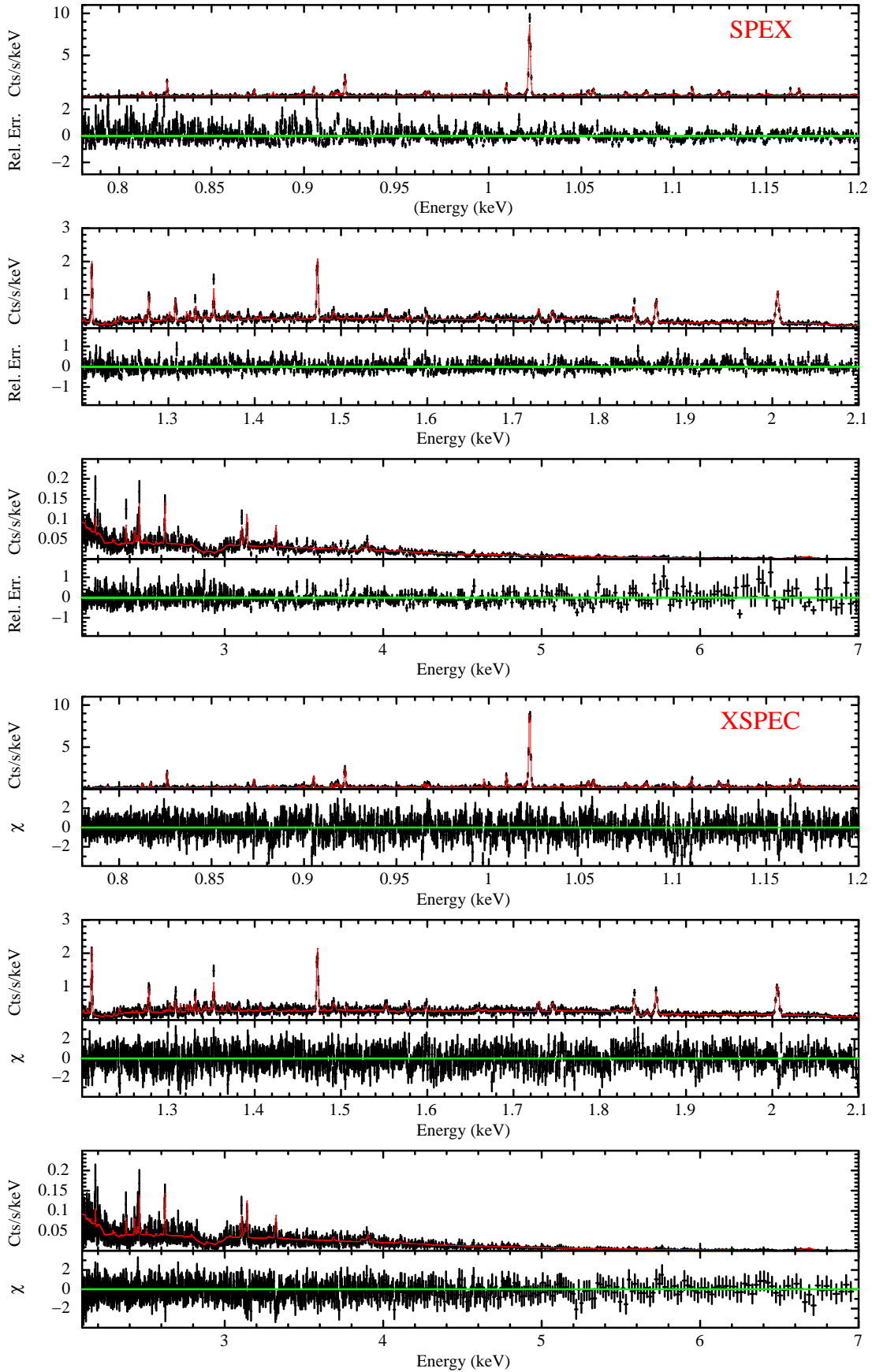

  \centering
  \includegraphics[width=11.5cm,angle=-90]{hr1099_plot_spex_new}
  \includegraphics[width=11.5cm,angle=-90]{hr1099_plot_xspec_new}  
  \caption{\label{fig:hr1099_plot_spex} Plots of the average HEG +1 spectrum from HR\,1099 extracted from the ObsID~62538. The top figure shows the results obtained within the {\sc spex} environment, while the figure on the bottom corresponds to the {\sc xspec} results. For clarity, the spectra in each figure have been split in three energy bands to ease the visualization of the relevant emission lines. Red solid lines represent the best fit model (see Table~\ref{tab:hr1099_spe}). We also show in each case the residuals from the best fits in the bottom panels. Note that {\sc spex} residuals are calculated as (Data-Model)/Model, while {\sc xspec} residuals are calculated as (Data-Model)/Error.}   
\end{figure*}

 \begin{figure*}
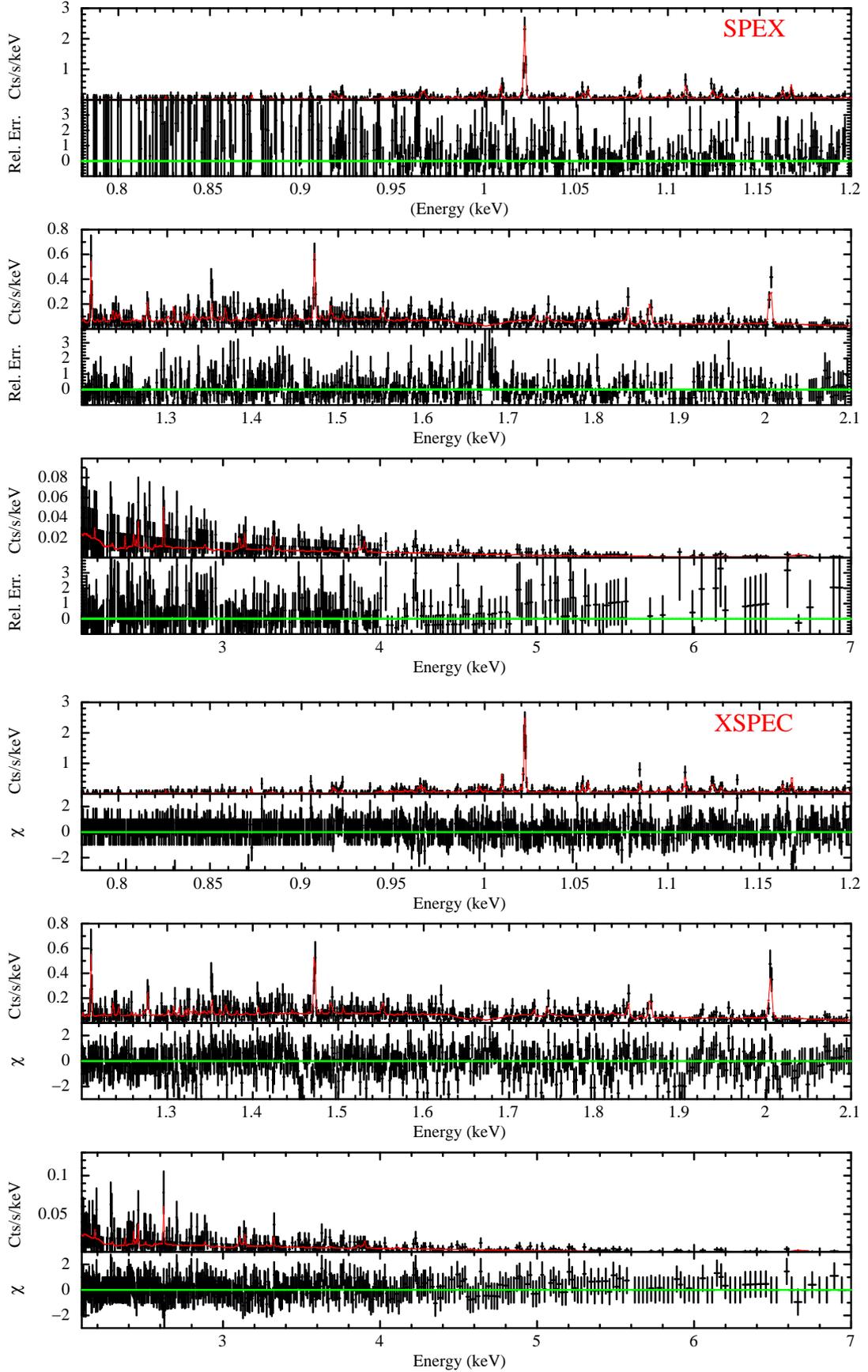

  \centering
  \includegraphics[width=11.5cm,angle=-90]{impeg_plot_spex_new}
  \includegraphics[width=11.5cm,angle=-90]{impeg_plot_xspec_new}  
  \caption{\label{fig:implot_plot_spex} Same as Fig.~\ref{fig:hr1099_plot_spex} but in the case of the HEG -1 average spectrum from IM\,Peg extracted from the ObsID~2527 (see also Table~\ref{tab:impeg_spe}).}   
\end{figure*}

 \begin{figure*}
  \centering
  \includegraphics[width=11.5cm,angle=-90]{arlac_plot_spex_new}
  \includegraphics[width=11.5cm,angle=-90]{arlac_plot_xspec_new}  
  \caption{\label{fig:arlac_plot_spex} Same as Fig.~\ref{fig:hr1099_plot_spex} but in the case of the HEG +1 average spectrum from AR\,Lac extracted from the ObsID~6 (see also Table~\ref{tab:arlac_spe}).}   
\end{figure*}

 \begin{figure*}
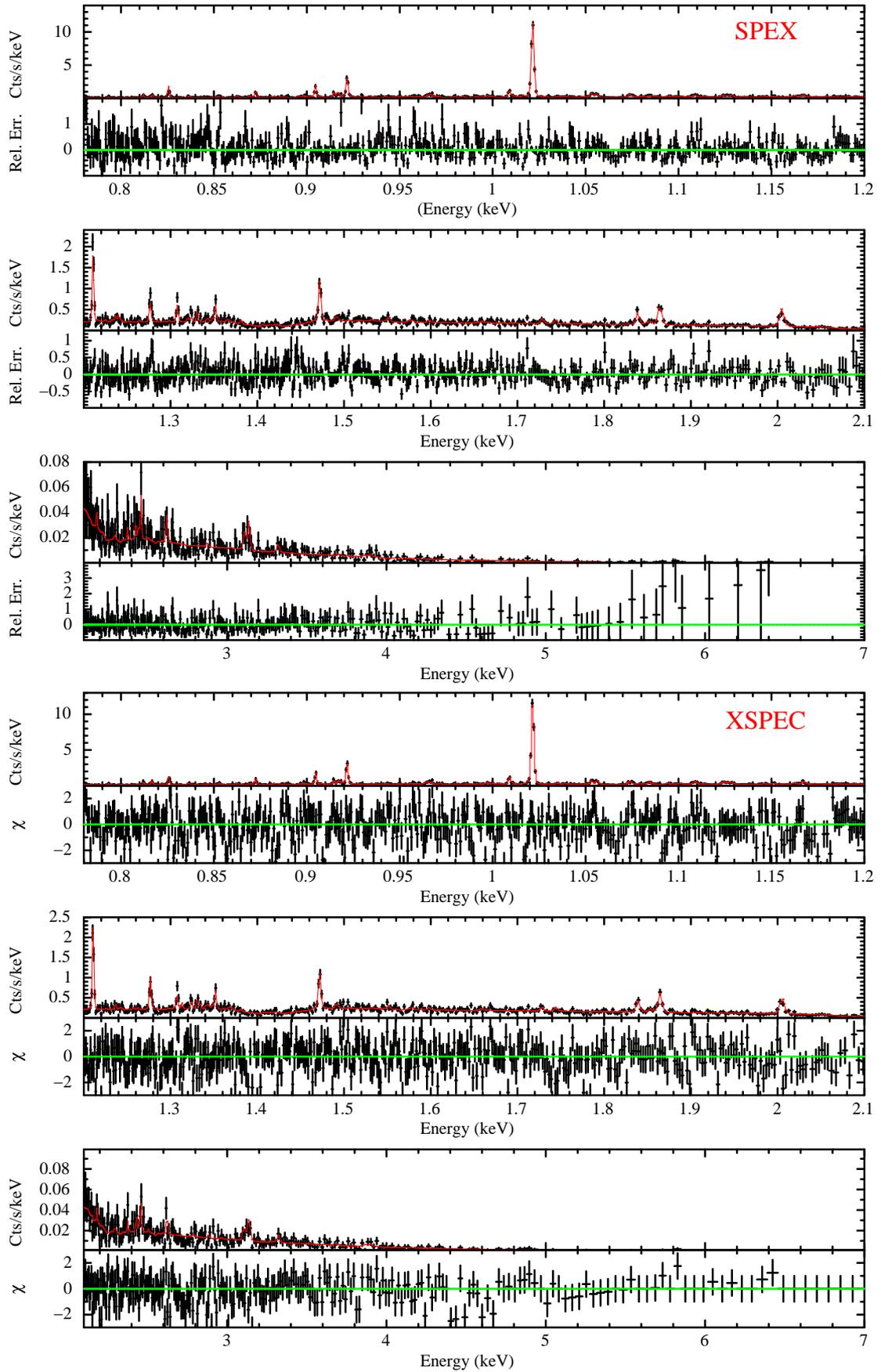

  \centering
  \includegraphics[width=11.5cm,angle=-90]{uxari_plot_spex_new}
  \includegraphics[width=11.5cm,angle=-90]{uxari_plot_xspec_new}  
  \caption{\label{fig:uxari_plot_spex} Same as Fig.~\ref{fig:hr1099_plot_spex} but in the case of the MEG +1 average spectrum from UX\,Ari extracted from the ObsID~605 (see also Table~\ref{tab:uxari_spe}).}   
\end{figure*}

 \begin{figure*}
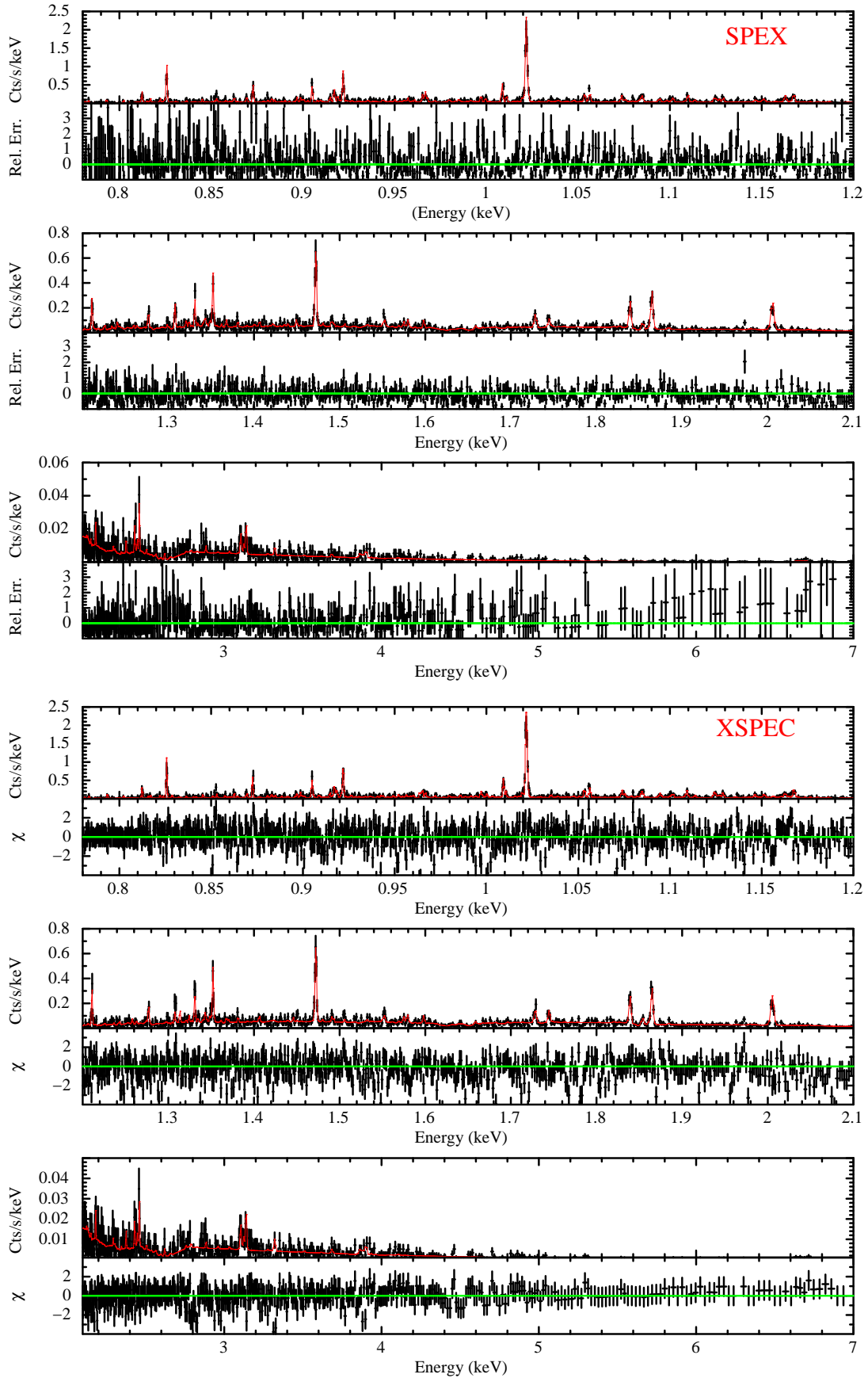

  \centering
  \includegraphics[width=11.5cm,angle=-90]{v824_plot_spex_new}
  \includegraphics[width=11.5cm,angle=-90]{v824_plot_xspec_new}  
  \caption{\label{fig:v824_plot_spex} Same as Fig.~\ref{fig:hr1099_plot_spex} but in the case of the HEG +1 average spectrum from V824\,Ara extracted from the ObsID~2538 (see also Table~\ref{tab:v824ara_spe}).}   
\end{figure*}

 \begin{figure*}
  \centering
  \includegraphics[width=11.5cm,angle=-90]{tzcrb_plot_spex_new}
  \includegraphics[width=11.5cm,angle=-90]{tzcrb_plot_xspec_new}  
  \caption{\label{fig:tzcrb_plot_spex} Same as Fig.~\ref{fig:hr1099_plot_spex} but in the case of the MEG +1 average spectrum from TZ\,Crb extracted from the ObsID~15 (see also Table~\ref{tab:tzcrb_spe}).}   
\end{figure*}

 \begin{figure*}
  \centering
  \includegraphics[width=11.5cm,angle=-90]{hr5110_plot_spex_new}
  \includegraphics[width=11.5cm,angle=-90]{hr5110_plot_xspec_new}  
  \caption{\label{fig:hr5110_plot_spex} Same as Fig.~\ref{fig:hr1099_plot_spex} but in the case of the MEG +1 average spectrum from HR\,5110 extracted from the ObsID~15 (see also Table~\ref{tab:hr5110_spe}).}   
\end{figure*}

 \begin{figure*}
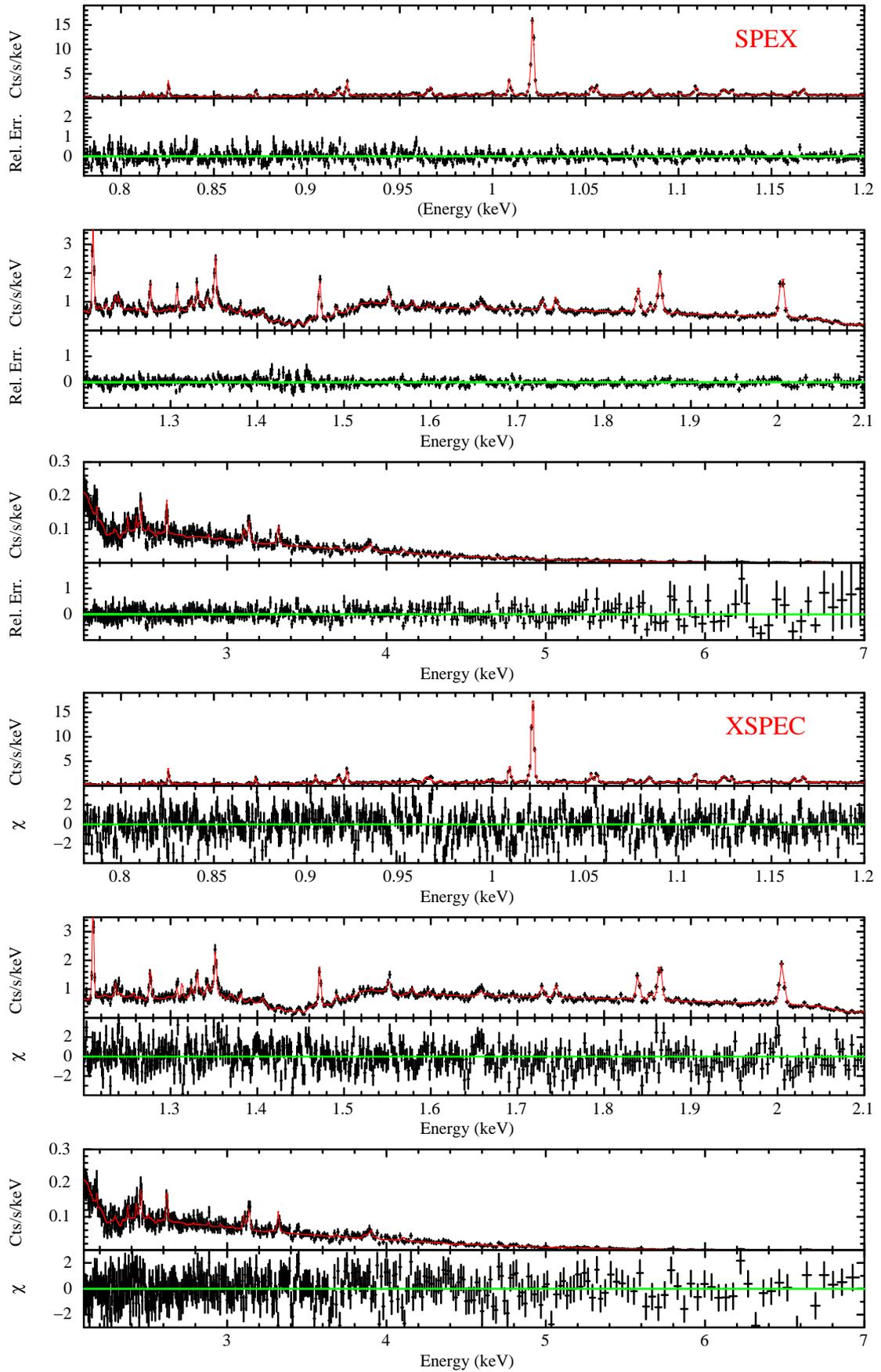

  \centering
  \includegraphics[width=11.5cm,angle=-90]{sigmagem_plot_spex_new}
  \includegraphics[width=11.5cm,angle=-90]{sigmagem_plot_xspec_new}  
  \caption{\label{fig:sigmagem_plot_spex} Same as Fig.~\ref{fig:hr1099_plot_spex} but in the case of the MEG +1 average spectrum from $\sigma$\,Gem  extracted from the ObsID~5422 (see also Table~\ref{tab:sigmagem_spe}).}   
\end{figure*}

 \begin{figure*}
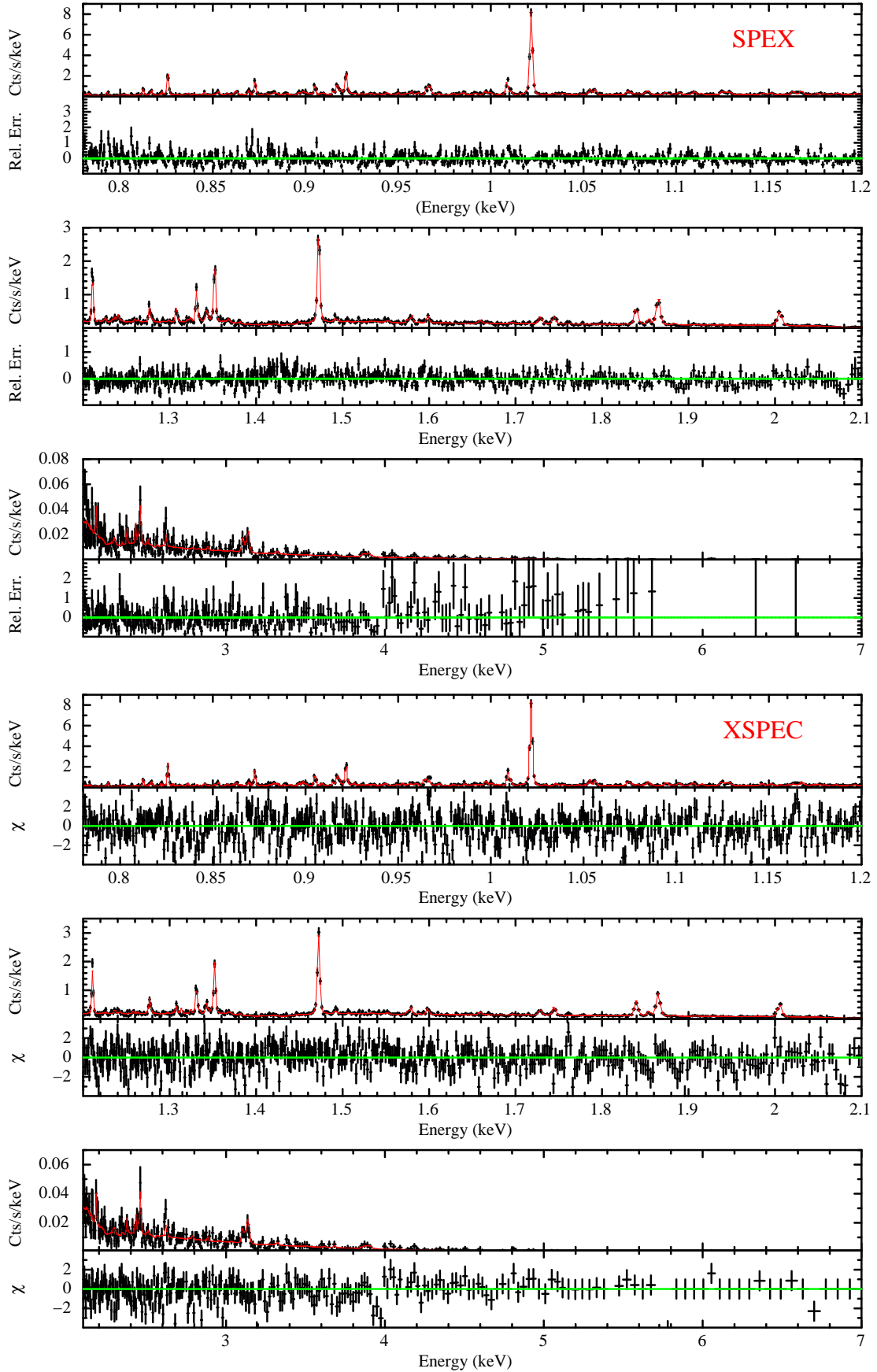

  \centering
  \includegraphics[width=11.5cm,angle=-90]{lambdaand_plot_spex_new}
  \includegraphics[width=11.5cm,angle=-90]{lambdaand_plot_xspec_new}  
  \caption{\label{fig:lambdaand_plot_spex} Same as Fig.~\ref{fig:hr1099_plot_spex} but in the case of the MEG +1 average spectrum from $\lambda$\,And  extracted from the ObsID~609 (see also Table~\ref{tab:lambdaand_spe}).}    
\end{figure*}

 \begin{figure*}
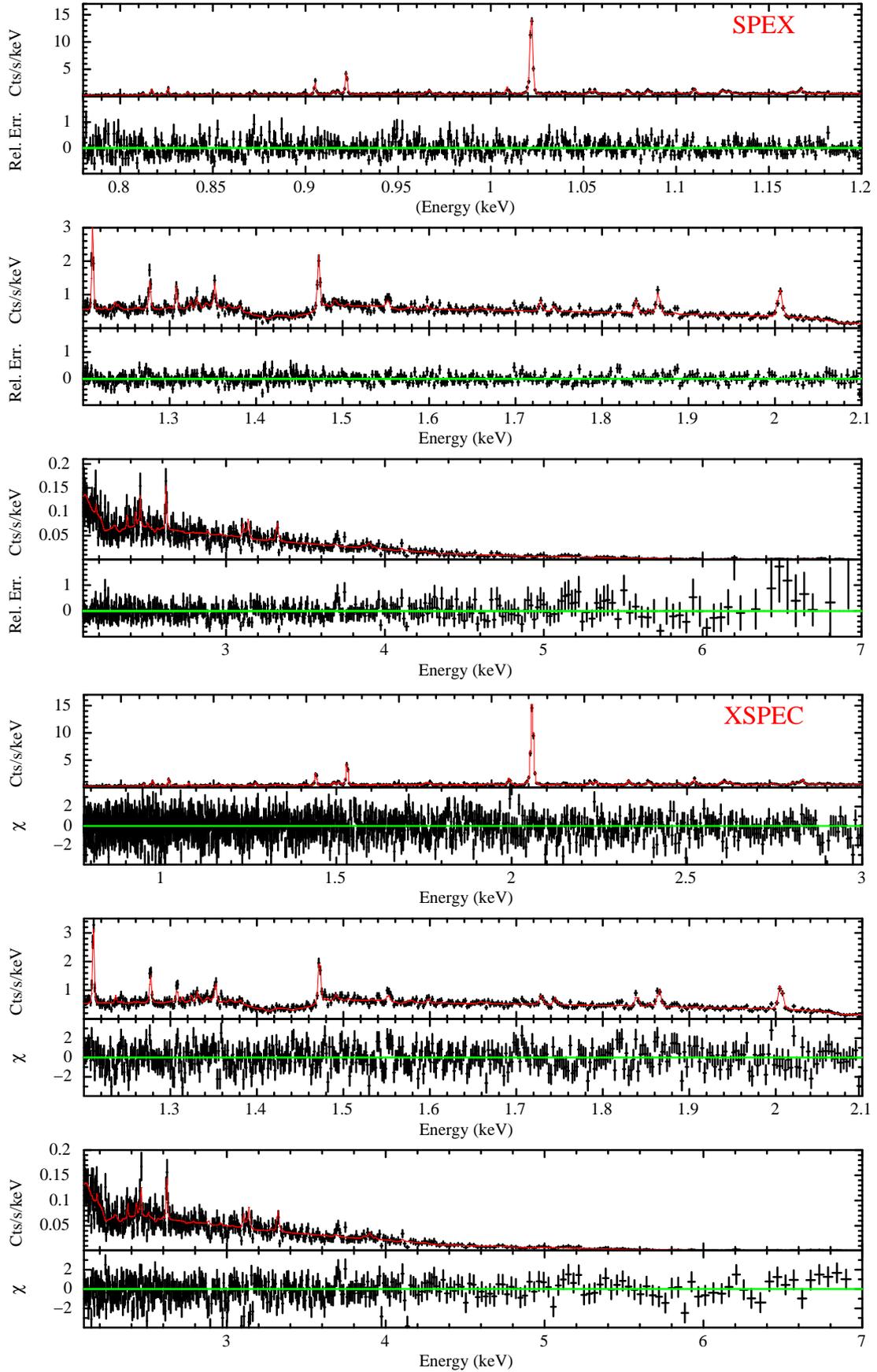

  \centering
  \includegraphics[width=11.5cm,angle=-90]{iipeg_plot_spex_new}
  \includegraphics[width=11.5cm,angle=-90]{iipeg_plot_xspec_new}  
  \caption{\label{fig:iipeg_plot_spex} Same as Fig.~\ref{fig:hr1099_plot_spex} but in the case of the MEG +1 average spectrum from II\,Peg extracted from the ObsID~1415 (see also Table~\ref{tab:lambdaand_spe}).}    
\end{figure*}

 \begin{figure*}
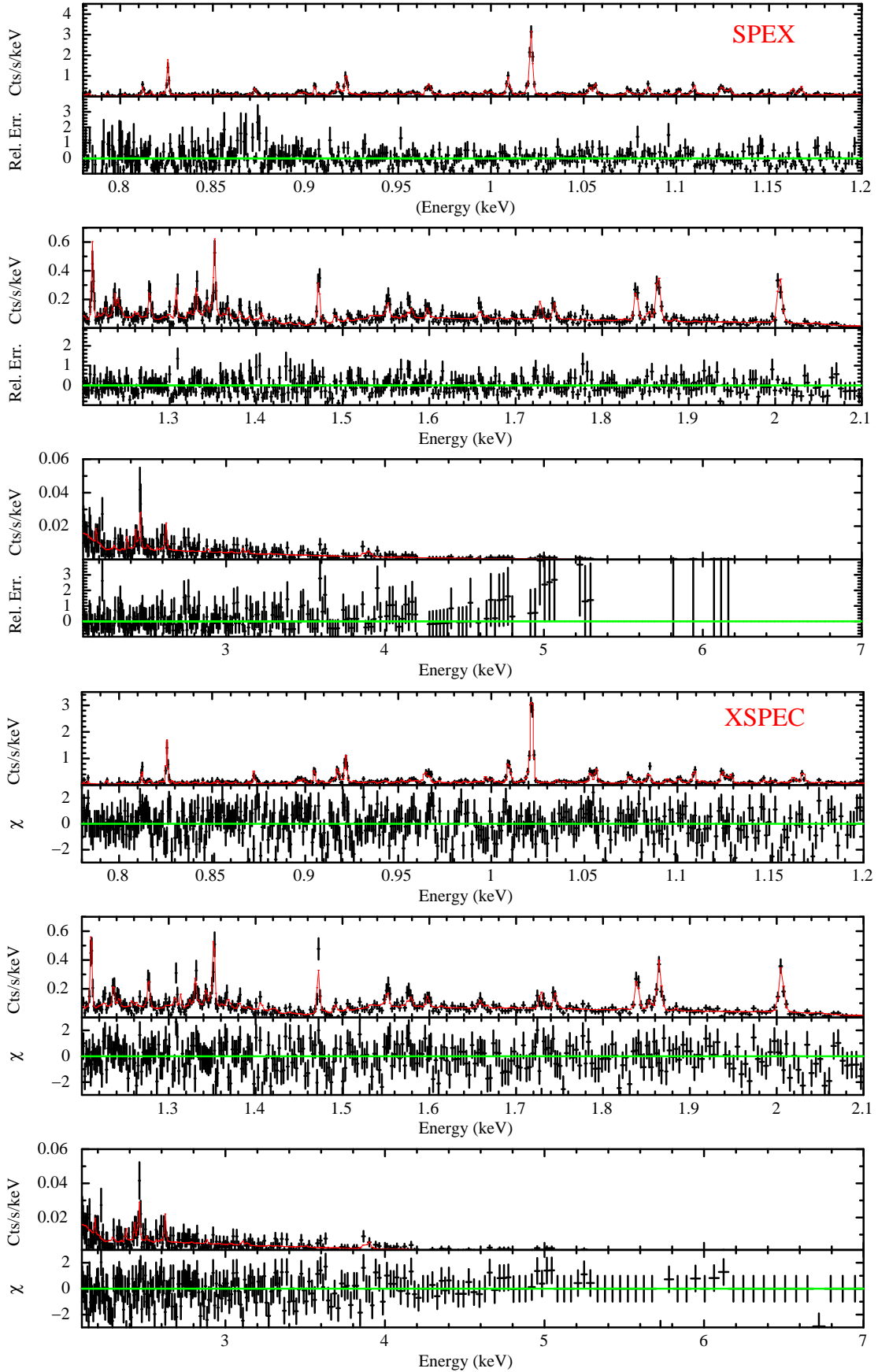

  \centering
  \includegraphics[width=11.5cm,angle=-90]{typyx_plot_spex_new}
  \includegraphics[width=11.5cm,angle=-90]{typyx_plot_xspec_new}  
  \caption{\label{fig:typyx_plot_spex} Same as Fig.~\ref{fig:hr1099_plot_spex} but in the case of the MEG +1 average spectrum from Ty\,Pyx extracted from the ObsID~601 (see also Table~\ref{tab:lambdaand_spe}).}    
\end{figure*}

\label{lastpage}
\end{document}